\documentclass[a4paper]{article}
\usepackage{amsmath,amssymb,bm,cellspace,graphicx,longtable,rotating,color,array,delarray,rotating,colordvi}

\pdfoutput=1
\newcommand{\dd}{\mathrm{d}}
\newcommand{\D}{}
\newcommand{\tc}[2]{#2}
\definecolor{darkgreen}{rgb}{0,0.5,0}
\definecolor{darkred}{rgb}{0.5,0,0}
\begin{document}
\title{
Random numbers from the tails of probability distributions using the 
transformation method}
\author{\normalsize \textsc{Daniel Fulger}$^{1,2,3}$,
\textsc{Enrico Scalas}$^1$ and \textsc{Guido Germano}$^3$
\medskip \\
\normalsize $^1$ Complex Systems Laboratory\\
\normalsize Department of Advanced Sciences and Technology\\
\normalsize Amedeo Avogadro University of East Piedmont\\
\normalsize Via Vincenzo Bellini 25 G, 15100 Alessandria, Italy\\
\normalsize www.mfn.unipmn.it/$\sim$scalas
\smallskip \\
\normalsize $^2$ Complex Systems Lagrange Laboratory\\
\normalsize Institute for Scientific Interchange\\ 
\normalsize Viale Settimio Severo 65, 10133 Torino, Italy
\smallskip \\
\normalsize $^3$ Computer Simulation Group, FB 15\\
\normalsize Philipps-University Marburg, 35032 Marburg, Germany\\
\normalsize www.staff.uni-marburg.de/$\sim$germano}
\maketitle

\begin{abstract}
The speed of many one-line transformation methods for the production of,
for example, L\'evy alpha-stable random numbers, which generalize Gaussian ones,
and Mittag-Leffler random numbers, which generalize exponential ones,
is very high and satisfactory for most purposes.
However, for the class of  decreasing probability densities fast
rejection implementations like the Ziggurat by Marsaglia and
Tsang promise a significant speed-up if it is possible to complement them 
with a method that samples the tails of the infinite support.
This requires the fast generation of random numbers
greater or smaller than a certain value. We present a method to
achieve this, and also to generate random numbers within any arbitrary interval.
We demonstrate the method showing the properties of the transform maps of the
above mentioned distributions as examples of stable and geometric stable
random numbers used for the stochastic solution of the space-time fractional diffusion equation.
\end{abstract}

\date{\today}

\maketitle

\section{Introduction}

\begin{figure}
\begin{center}
\includegraphics[clip=true,height=9.5cm,angle=-90]{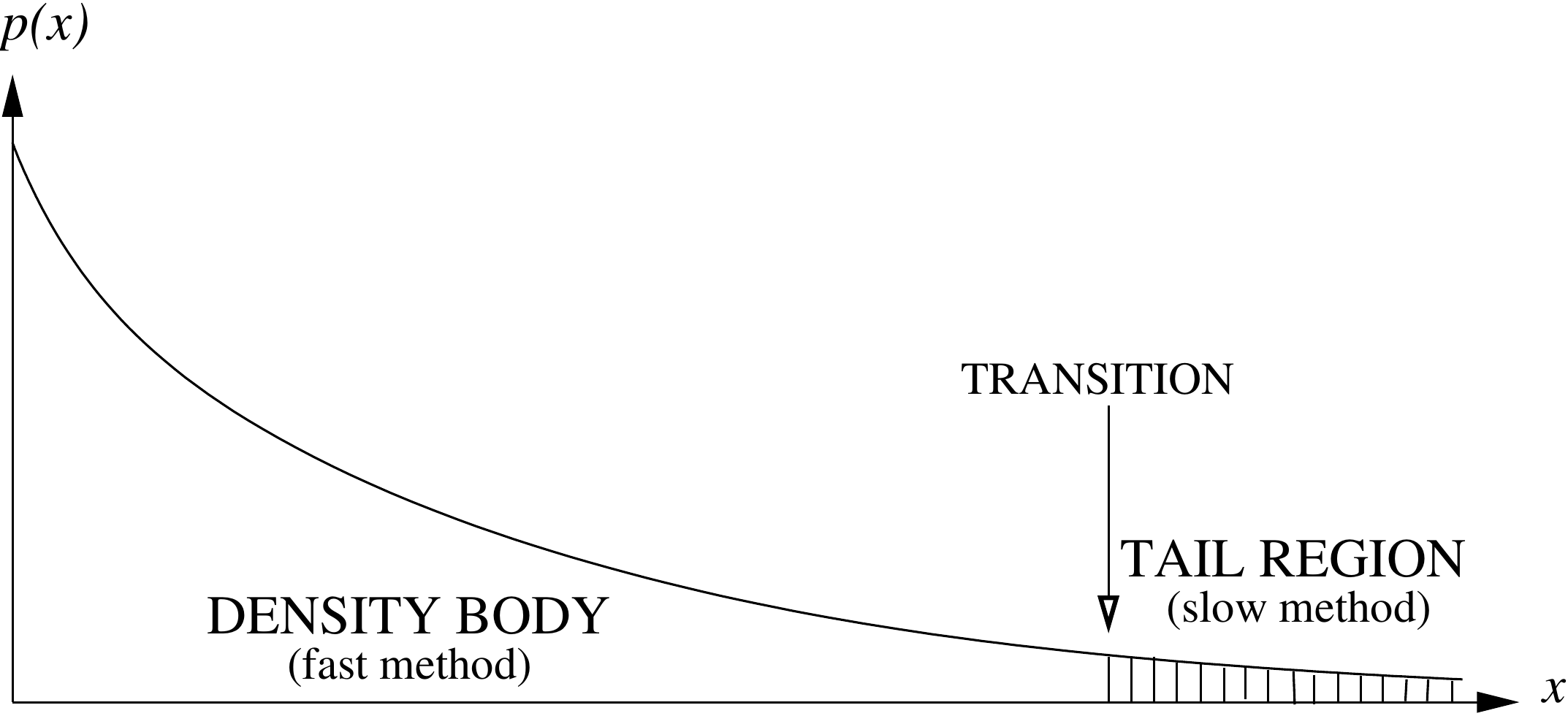}
\end{center}
\caption{ \label{fig:schema} Schematic illustration of using two methods
for sampling a distribution. A fast method may be available only for the body,
while the tails can be sampled with a slower method.}
\end{figure}

Many numerical methods for the generation of random numbers 
represent   the main body of the probability density using a fast method and the
tails using an alternative method. A famous
example is the Ziggurat method by Marsaglia and Tsang~\cite{Marsaglia1984}.  
Fig.~\ref{fig:schema} depicts the situation schematically.
A reason for this apparent complication is that the method for the main body works
best and fastest on a finite support or is specially designed for the
main body in terms of accuracy or speed. Handling the tails efficiently is
often more involved,  especially with difficult non-invertible
densities with infinite support. Since rarely needed, variates
from the tail can safely be generated by a slower
method~\cite{Devroye1989,Marsaglia1984,Marsaglia2000}.  Overall, a significant
speed-up can be achieved. In this paper
we show how  to sample directly and efficiently via a rejection
technique a random number  $X$ such that  $X>x$ or $X<x$ where
$x\in(-\infty, +\infty)$ at least within the limits of the  numerical
representation. This is achieved by using properties of the transform representation
of the distributions.  The examples we use for demonstration are the L\'evy
$\alpha$-stable ~\cite{Levy1925,Nolan1997,Nolan1999} and the Mittag-Leffler 
one-parameter probability densities~\cite{Hilfer2006}. 
A transformation formula for the former is well known~\cite{Chambers1976,Weron2004},  
while the transform representation of the latter was
discovered~\cite{Devroye1996,Jayakumar2003,Kozubowski1998,Kozubowski2000,Kozubowski2001,
Kozubowski1999,Pakes1998} and applied~\cite{Germano2006,Fulger2008,Germano2008} only
recently. The two distributions are generalizations of the Gaussian and exponential distribution
respectively and play an important role together for the
solution of the space-time fractional diffusion equation.

Our rejection concept is general for any distribution
that provides a transform representation. It can  sample efficiently from
arbitrary infinite or finite intervals as opposed to other existing  methods that
are designed especially for certain densities. In this work we do
not consider the technical details of a speed-optimized implementation,
but  explain the basis of the algorithm and show  example
applications. The method is based on properties of the
two-dimensional transform maps that seem unnoticed yet.

The assumption for using the method introduced here is that the tail region
requires high accuracy due to high demands on statistics as well as speed.
The transformation formula by Chambers, Mallows and  Stuck~\cite{Chambers1976}
for example is exact and for most applications the recommended method  for the production
of L\'evy $\alpha$-stable random numbers~\cite{Weron2004}.
The replacement of the tails by a simple invertible Pareto function
is not totally  appropriate because this is only an asymptotic approximation; 
moreover it introduces a transition region. The
more sophisticated and smooth  this transition, the more complicated
and slower the overall procedure.  Such a replacement of the tail
contrasts the initial goal of speed.  But the most demanding
contemporary applications of random numbers~\cite{Meerschaert2002,Weron2004},
as of the two suitable examples we treat here, will require large amounts and
therefore fast production.  The tails should be accurate without 
an approximated  transition region from the 
density body to its tails.  In some cases fast series
expansion methods can be used but  with a
compromise in {accuracy~\cite{Devroye1989}}. A more detailed analysis of such
considerations can be found in Ref.~\cite{Thomas2007b}, where most known
algorithms for the Gaussian distribution (as a simple and special case
of the L\'evy $\alpha$-stable distribution) are analyzed in the context of
contemporary statistical applications as well as expectations of future
demands.  It is argued extensively how speed of production implies
the demand for very many random numbers, which in turn requires greater
accuracy of the resulting distribution.

Consider  the Ziggurat rejection
method by  Marsaglia et al.~\cite{Marsaglia1964b,Marsaglia1964a, Marsaglia1984} that was introduced to
produce Gaussian and exponential random numbers. It is an exact method up to the
numerical limits  of floating point representation.  In principle it
is applicable to all decreasing or symmetric densities, 
provided a suitable tail sampling
method is available~\cite{Marsaglia1964aa}. In particular the  implementation by Marsaglia and
Tsang~\cite{Marsaglia2000} and a recent
version by Rubin and Johnson~\cite{Rubin2006} are 
about two orders of magnitude faster on  contemporary processors 
than other dedicated methods for Gaussian and
exponential random variates; therefore it is likely
to outrun any non-trivial transformation method by at least the same factor. 
The hurdles to apply the Ziggurat method to other densities with infinite support,
with additional parameters and for which no closed  form or simple
transform exist are: a) the costly setup of the look-up table, b) the
necessity of equal areas of the rectangles covering the density as
well as the area under the tail and finally c) a reasonably 
fast and accurate tail sampling method. Difficulty a) must be evaluated  in
relation to the required number of variates if it is possible to predict
the  setup costs as a function of the density parameters.  
The meaning of ``fast'' in c) is defined by the ratio of tail variates versus
body variates and the speed of the body sampling method. A slow tail sampling can always be 
balanced by sufficiently infrequent calls to the latter.

Provided the complementary cumulative density function  $\int_x^\infty f(x')\, \dd x'$
can be computed sufficiently exact on demand, then any required value
of the tail surface, and thus any relative frequency of calls to the tail sampling function,
can be achieved  in the setup of the Ziggurat by
an iterative process. For the details of the setup
refer to Ref.~\cite{Marsaglia2000} \tc{darkred}{and for alternative 
concepts to Ref.~\cite{Rubin2006}.} Independently of such
considerations the production of  L\'evy $\alpha$-stable random
numbers in the tails, but also in arbitrary finite intervals, are
themselves  examples where the method introduced in this paper is
suitable. Of course the Ziggurat method is
applicable to non-symmetric decreasing densities by representing 
two halves  with separate generators which have to be called
alternatingly in a ratio that corresponds to the ratio of respective
areas covered by each halves. 
 
In Sec.~\ref{sec:levymap} we introduce the
L\'evy $\alpha$-stable probability density on the basis of which 
Sec.~\ref{sec:sampling} explains our method. In Sec.~\ref{sec:ML} the Mittag-Leffler distribution, 
its transform representation and transform map are presented.

\section{The L\'evy $\alpha$-stable probability density and its transform map}\label{sec:levymap}

A convenient representation of the  L\'evy  probability density function
in its most popular  parametrization~\cite{Nolan1997,Nolan1999, Weron2004}
is via the inverse Fourier transform of its characteristic function:
\begin{equation} \label{eqn:f_LevybyFT}
L_{ \alpha\beta\gamma\delta}(x)=\frac{1}{2\pi}\int_{-\infty}^\infty\:  
                        \phi_{ \alpha\beta\gamma\delta}(k)\: \exp(-ikx)\: \dd k,
\end{equation}
where
\begin{equation} \label{eqn:Levycharf}
\log \phi_{ \alpha\beta\gamma\delta}(k)
= \left\{
\begin{array}{ll}
-\gamma^\alpha|k|^\alpha\left(1 - i\beta\mbox{sign}(k)
\tan(\frac{2}{\pi}\alpha)\right)+i\delta k & \mbox{for}\quad \alpha\ne 1\\
& \\
-\gamma |k| \left(1 + i\beta\mbox{sign}(k) \frac{2}{\pi}
\log|k|\right)+i\delta k & \mbox{for} \quad \alpha= 1 .
\end{array}
\right.
\end{equation}
The index or order $\alpha\in(0,2]$ determines the exponent of the power-law
tail. The parameter $\beta\in[-1,1]$ governs the skewness,
$\gamma\in(0,\infty)$ the horizontal scale and $\delta\in(-\infty,\infty)$ the
location. The advantage of this parametrization is that the density 
and the distribution function are jointly continuous in all four parameters;
the same applies to the convergence to the power-law tail.
The last two parameters can safely be set to 1 and 0 without loss of generality.  
\tc{darkred}{Other values can be obtained through
\begin{equation}
X_{\alpha\beta\gamma\delta} = \gamma X_{\alpha\beta 1 0} + \delta  .
\end{equation}  }
We therefore omit $\gamma$ and $ \delta$ in the subscripts
and also $\beta$ if equal to zero. 
The symmetric  case with $\beta=0$ has the simpler form
of an inverse cosine transformation
\begin{equation} \label{eqn:f_Levysym}
L_{\alpha }(x)=\frac{1}{\pi}\int_0^\infty \exp({- k^\alpha})\,\cos(kx)\, \dd k .
\end{equation}
Rejection methods for L\'evy $\alpha$-stable random numbers that use 
asymptotic series representations of the  density function
are sometimes used if speed has highest priority~\cite{Devroye1989}.
However, the known types of series expansions for the L\'evy density
tend to become inaccurate especially in the tails and
also account for a certain fraction of
uniform random numbers to be lost (rejected) in the sampling. To achieve
best performance (minimum rejection rate and maximum accuracy) one must use different versions of the 
algorithms and expansions depending on the combination of parameter values and their range.
This is  the case in particular for $\beta\ne0$.  
A review on these methods and their deficiencies can be found in Ref.~\cite{Devroye1989}.

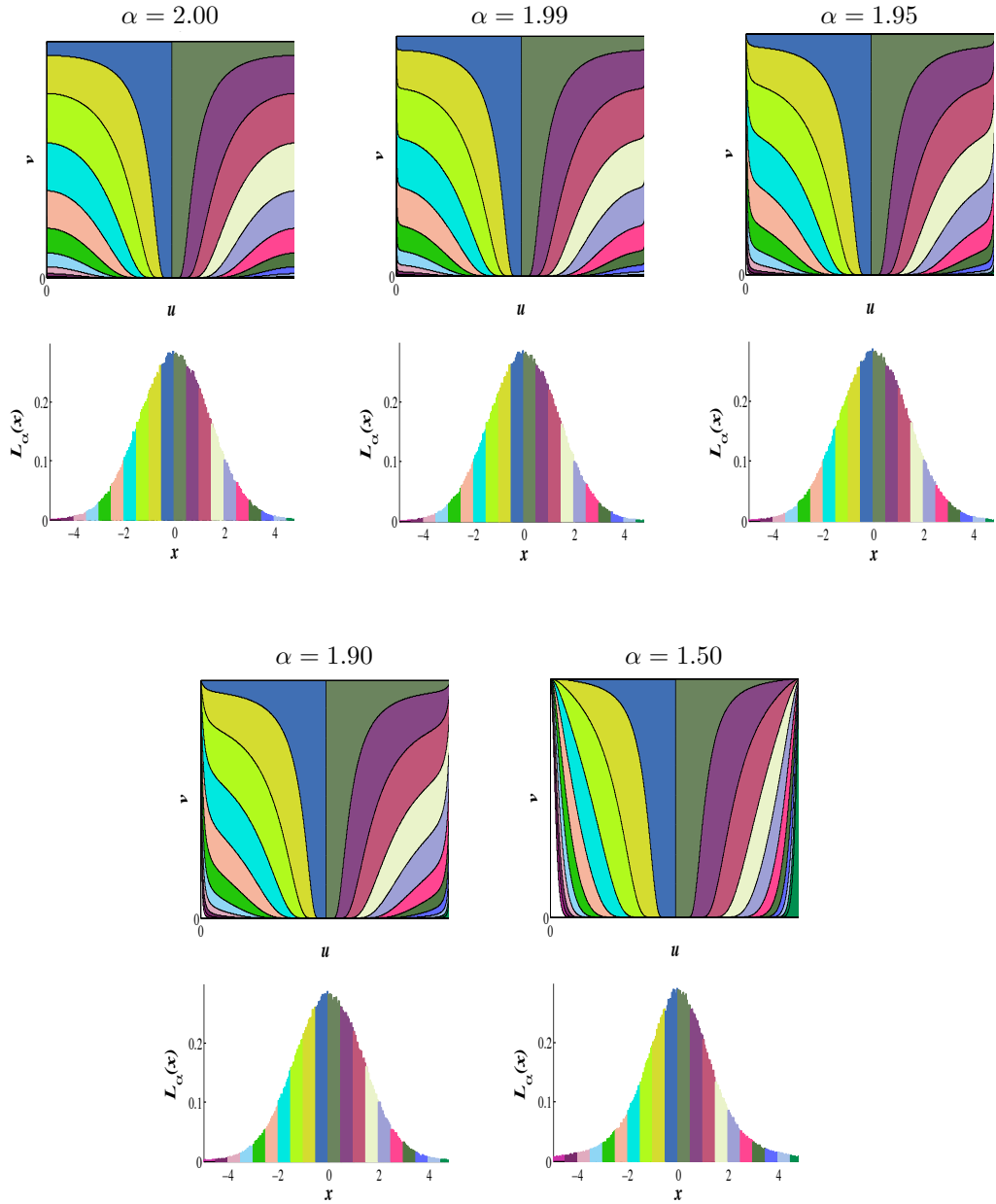
\begin{figure}
\setlength{\unitlength}{1cm}
\parbox{1.19\textwidth}{
\hspace{-0.7cm}\includegraphics[clip=true,height=8cm,width=4.5cm]{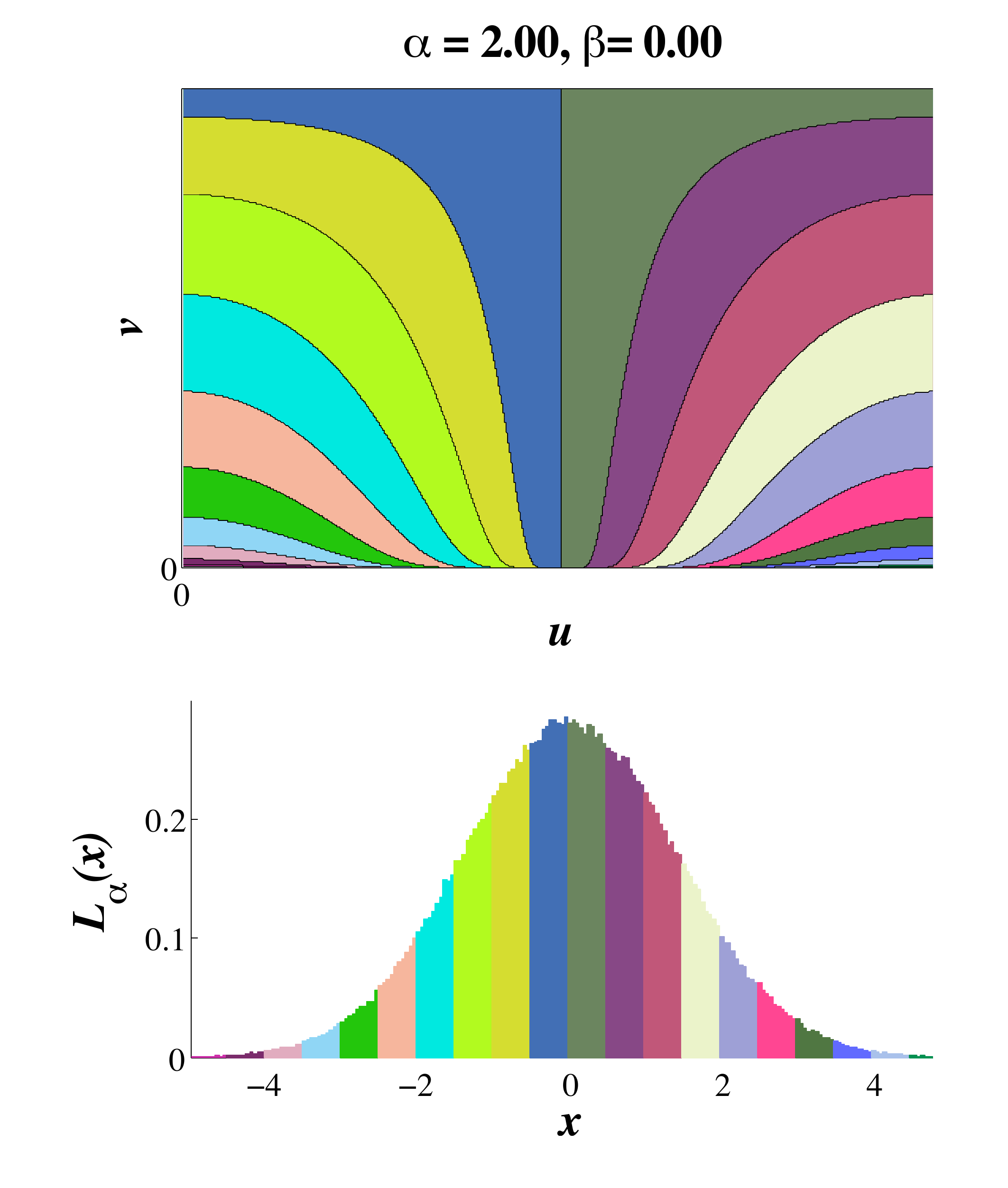}
\parbox{0cm}{\begin{picture}(0,0)(0,0)\put(-2.9,7.59){\parbox{0cm}{\colorbox{white}{$\alpha=2.00$}} }\end{picture}}
\includegraphics[clip=true,height=8cm,width=4.5cm]{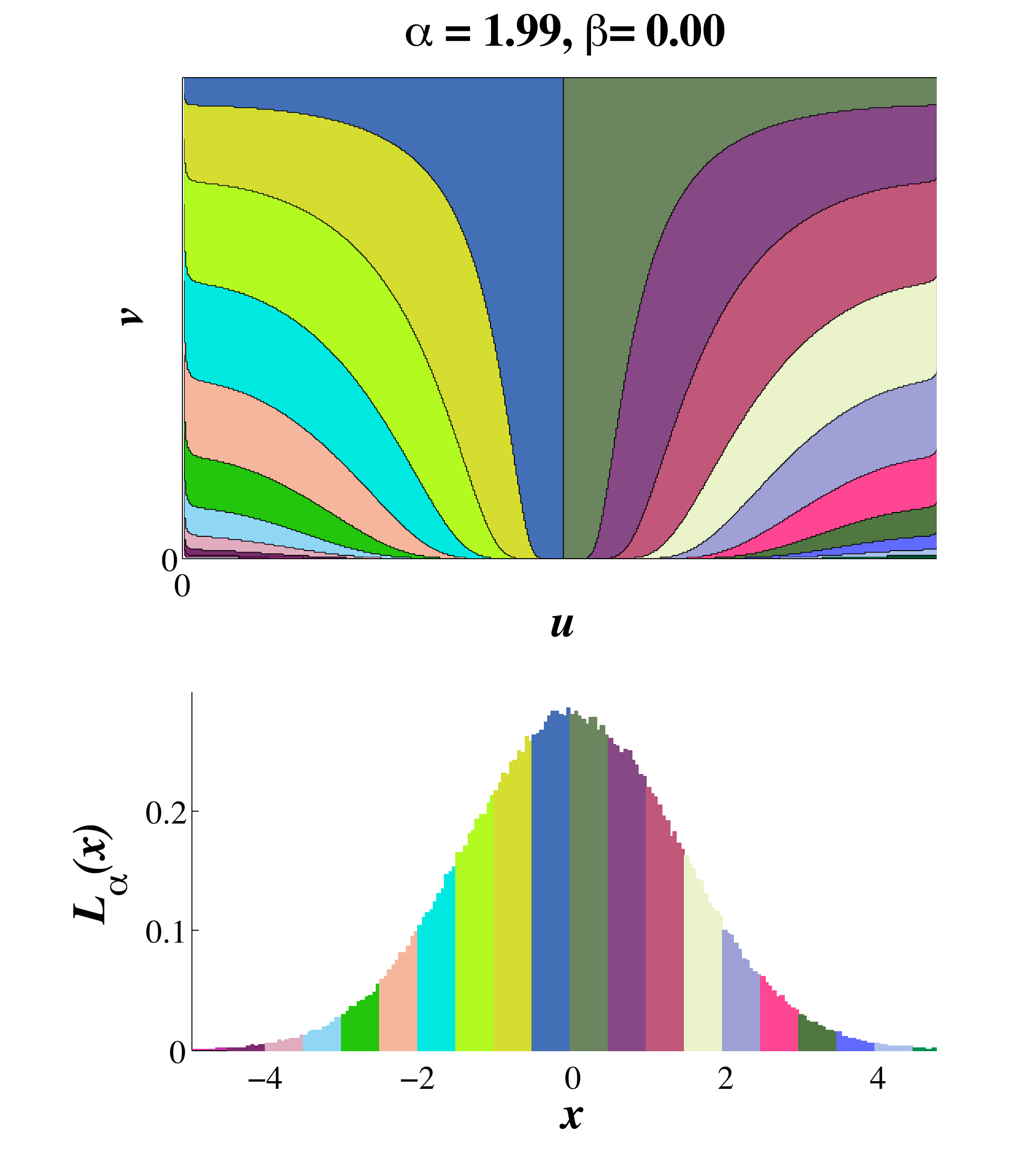}
\parbox{0cm}{\begin{picture}(0,0)(0,0)\put(-2.9,7.59){\parbox{0cm}{\colorbox{white}{$\alpha=1.99$}} }\end{picture}}
\includegraphics[clip=true,height=8cm,width=4.5cm]{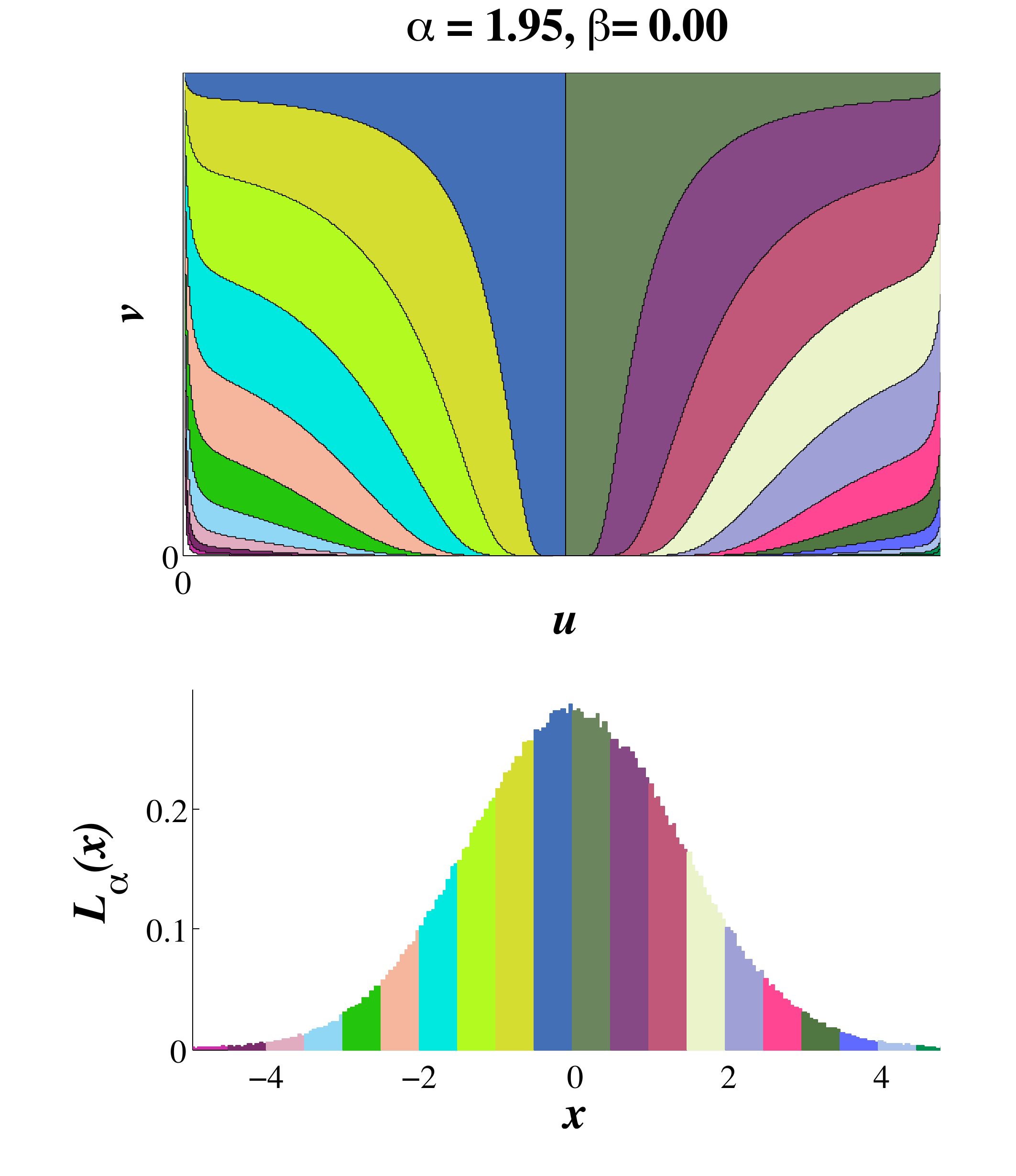}
\parbox{0cm}{\begin{picture}(0,0)(0,0)\put(-2.9,7.59){\parbox{0cm}{\colorbox{white}{$\alpha=1.95$}} }\end{picture}}\\
}
\begin{center}
\includegraphics[clip=true,height=8cm,width=4.5cm]{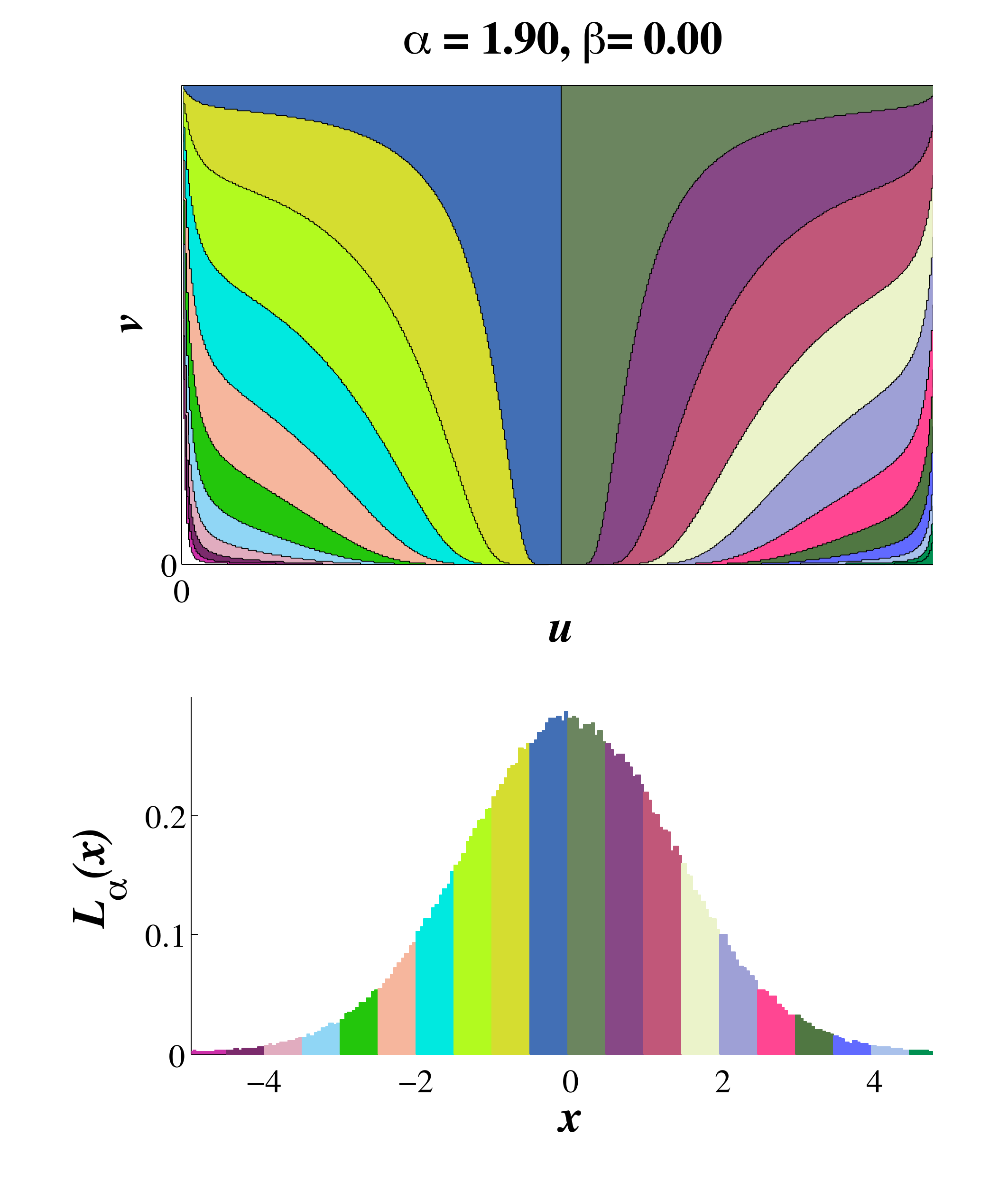}
\parbox{0cm}{\begin{picture}(0,0)(0,0)\put(-2.9,7.59){\parbox{0cm}{\colorbox{white}{$\alpha=1.90$}} }\end{picture}}
\includegraphics[clip=true,height=8cm,width=4.5cm]{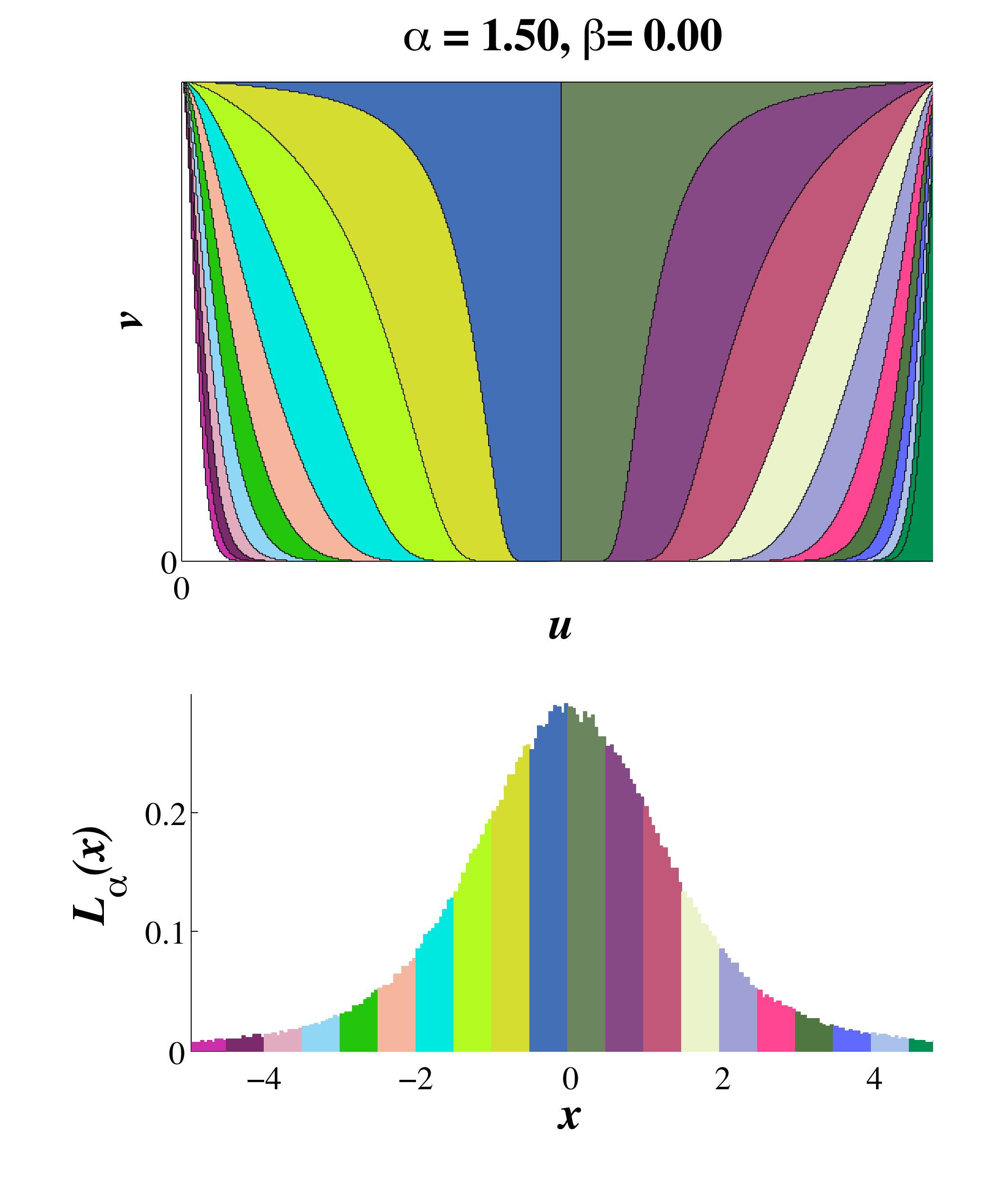}
\parbox{0cm}{\begin{picture}(0,0)(0,0)\put(-2.9,7.59){\parbox{0cm}{\colorbox{white}{$\alpha=1.50$}} }\end{picture}}\\
\end{center}
\caption{ \label{fig:mapsbet=0} (Color online) The map $X=F_{\alpha}(U,V)$ with 
$U,V\in(0,1)$
giving the symmetric L\'evy distribution $L_{\alpha}(x)$ for different values of $\alpha$. 
For $\alpha=2$ the picture corresponds to the
Box-Muller map for the generation of Gaussian 
random numbers. The bottom part of each map shows the respective histogram. Areas with equal
colors correspond to each other. Note that 
 the transition from $\alpha=2$ to  $\alpha<2$ is discontinuous for $u=0$ and $u=1$ 
and the points (0,1) and (1,1) develop a singularity. 
}
\end{figure}

\begin{figure}
\begin{center}
  \includegraphics[clip=true,height=8.1cm,width=4.9cm]{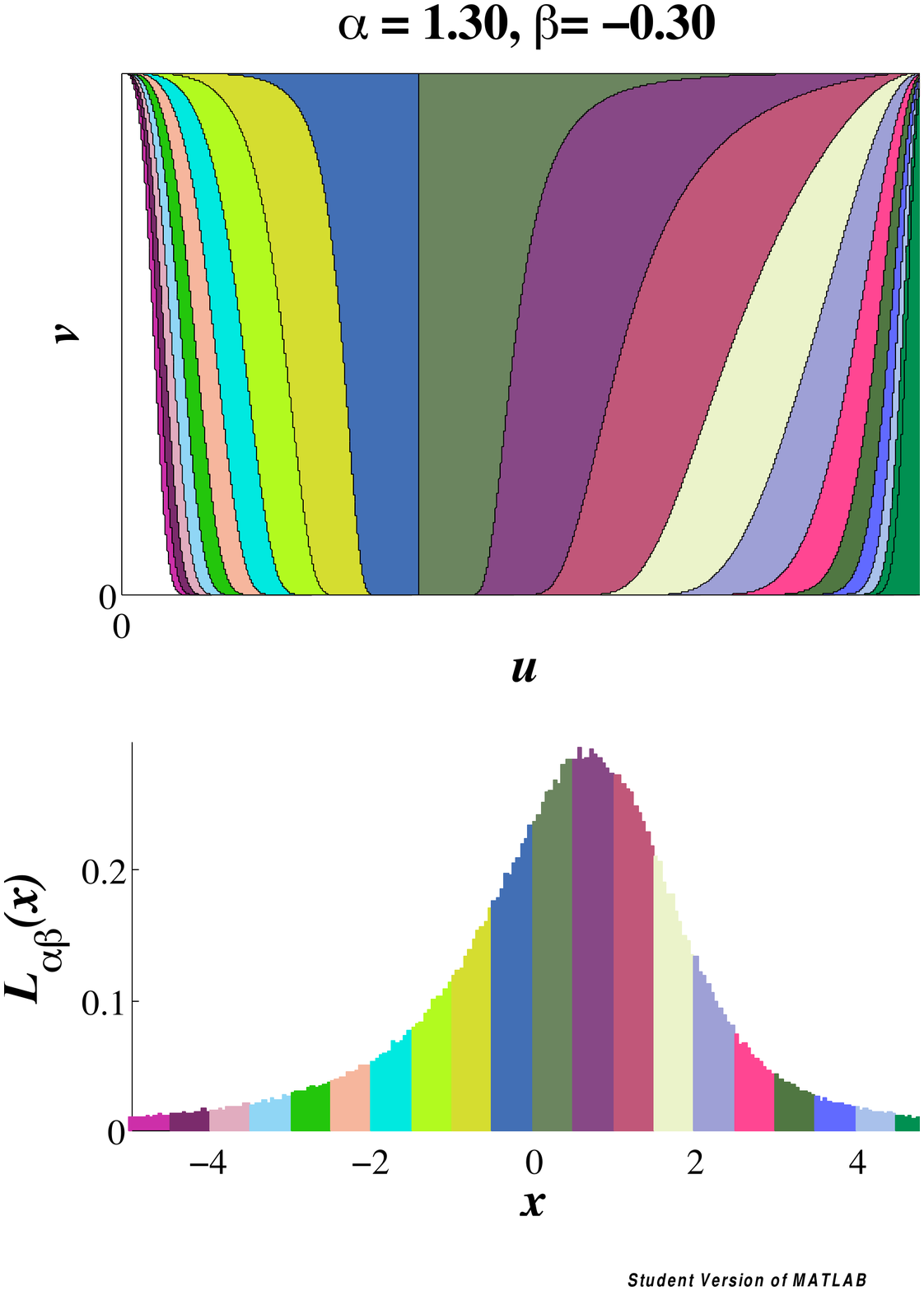}
  \includegraphics[clip=true,height=8cm,width=4.9cm]{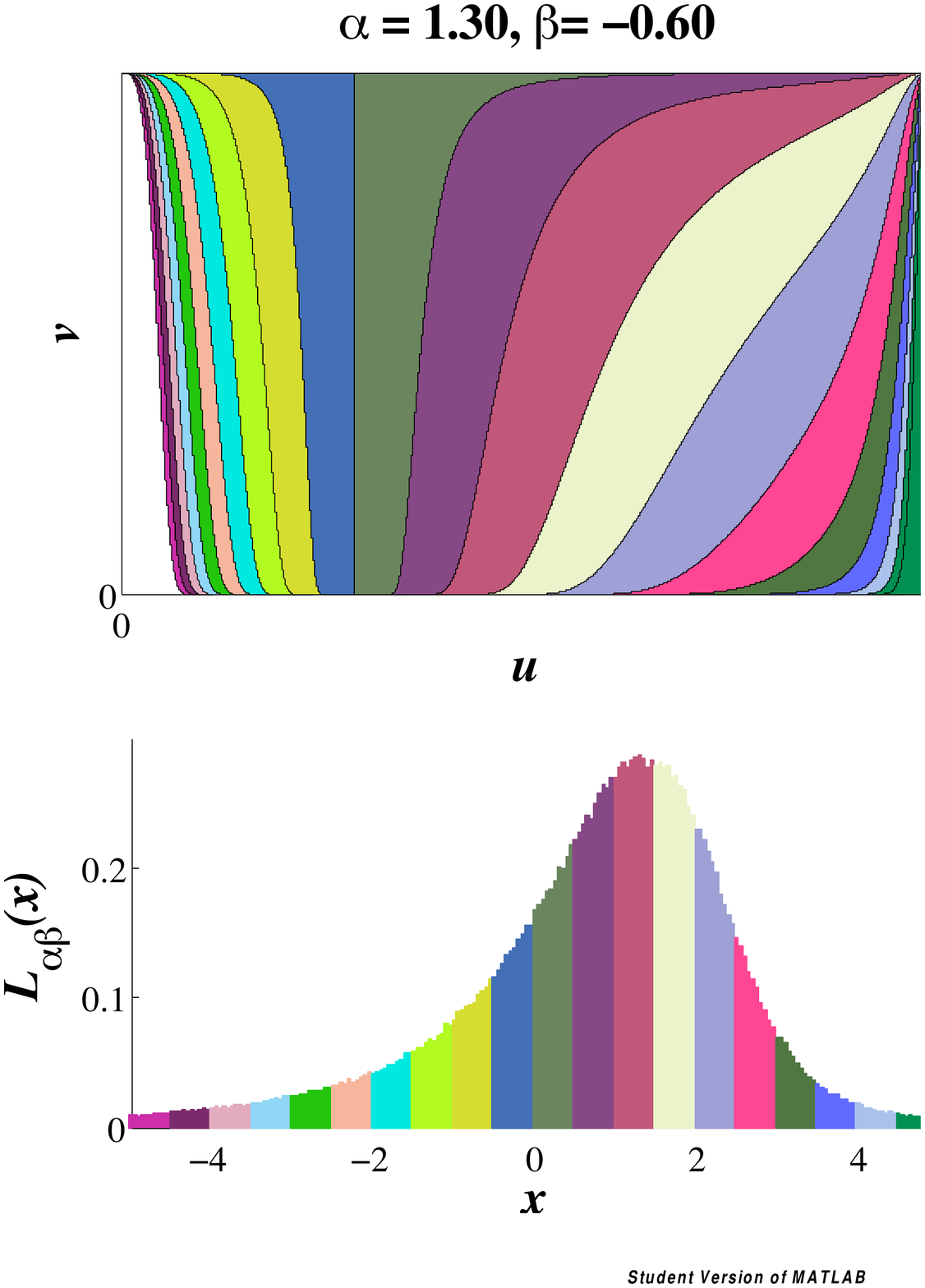}
\end{center}
\caption{ \label{fig:al130be-06} (Color online)  The map $X=F_{\alpha\beta}(u,v)$
giving the asymmetric L\'evy distribution $L_{\alpha\beta}(x)$ for 
two values of $\beta$. }
\end{figure}

\begin{figure}
  \setlength{\unitlength}{1cm}
\begin{center}
\scalebox{1.25}{
\hspace{-7mm}\parbox{\textwidth}{
\begin{picture}(0,0)(0,0)  \put(0.4,3){\scriptsize 1}\put(3.2,0.3){\scriptsize 1}\end{picture}
\includegraphics[clip=true,height=3.5cm,width=3.5cm]{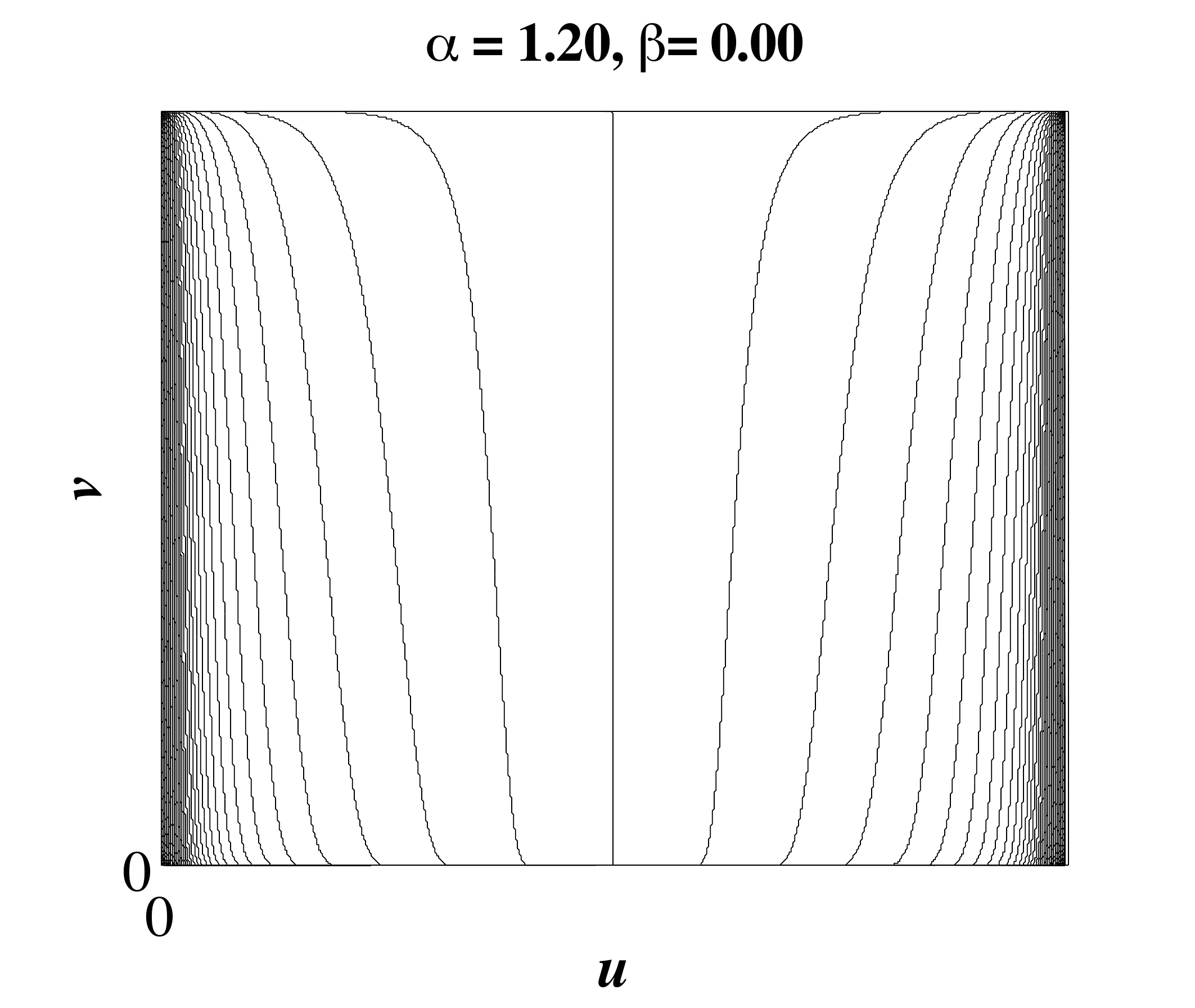}
\begin{picture}(0,0)(0,0)  \put(0.4,3){\scriptsize 1}\put(3.2,0.3){\scriptsize 1}\end{picture}
\includegraphics[clip=true,height=3.5cm,width=3.5cm]{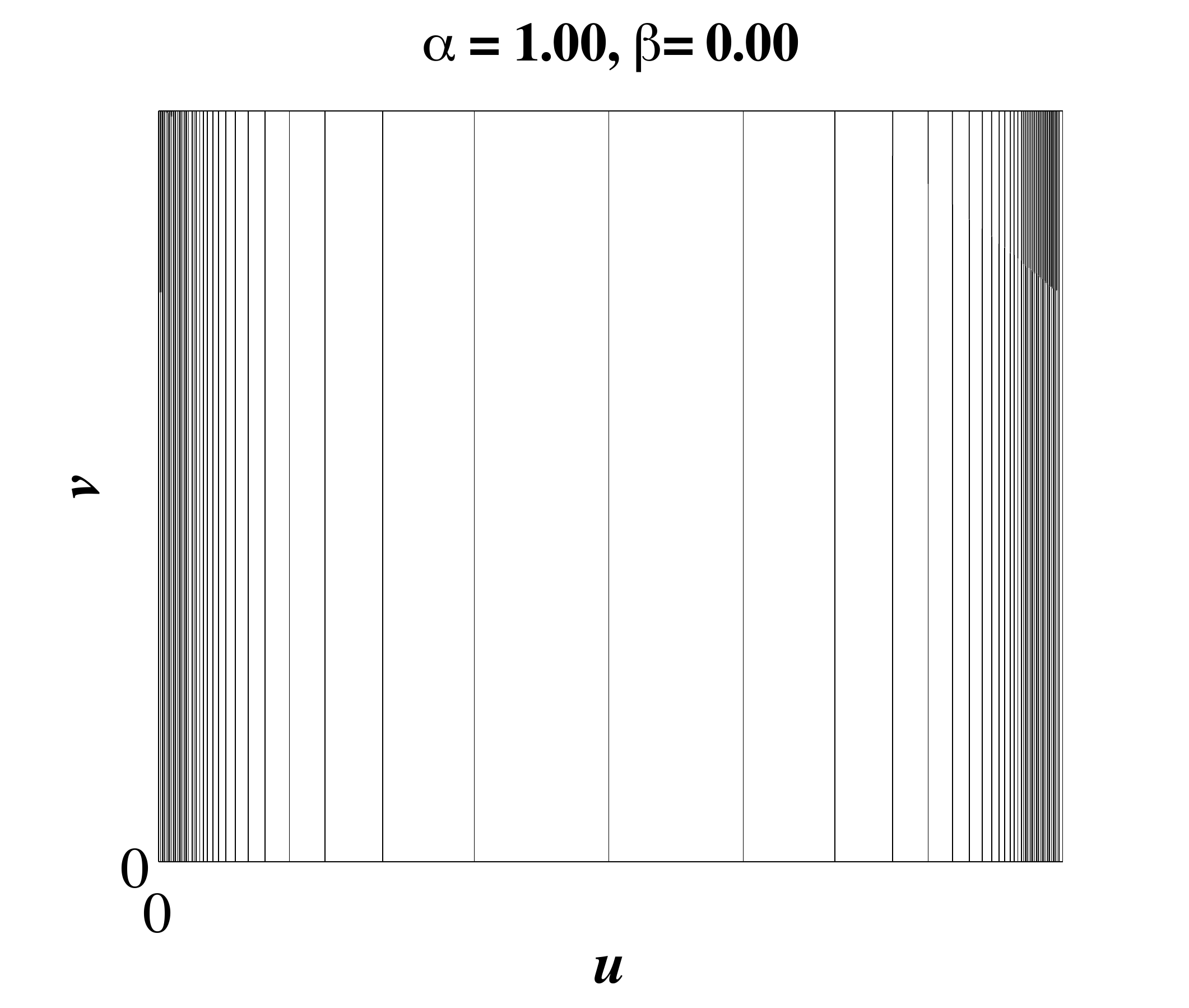}
\begin{picture}(0,0)(0,0)  \put(0.4,3){\scriptsize 1}\put(3.2,0.3){\scriptsize 1}\end{picture}
\includegraphics[clip=true,height=3.5cm,width=3.5cm]{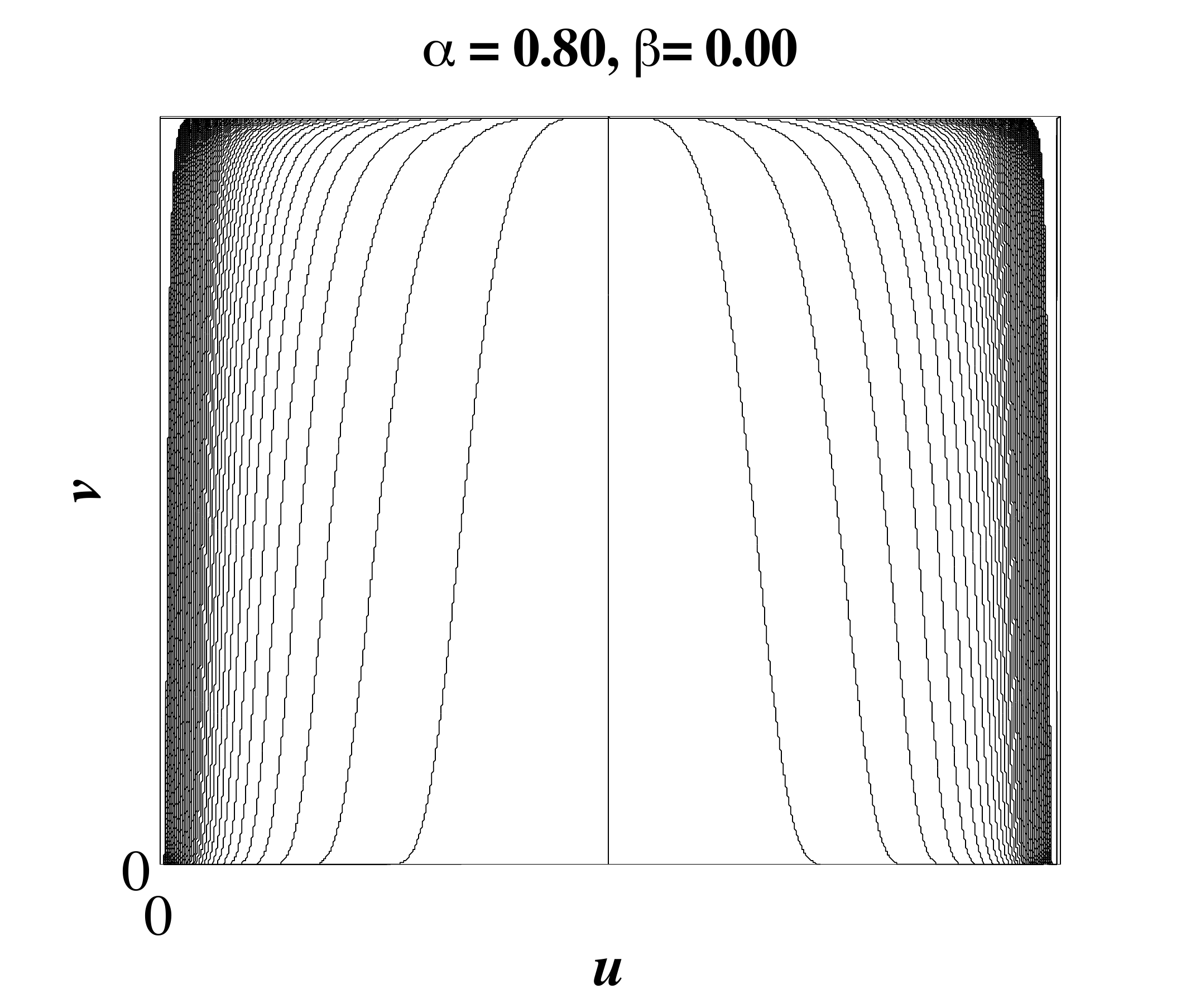}\\
\begin{picture}(0,0)(0,0)  \put(0.4,3){\scriptsize 1}\put(3.2,0.3){\scriptsize 1}\end{picture}
\includegraphics[clip=true,height=3.5cm,width=3.5cm]{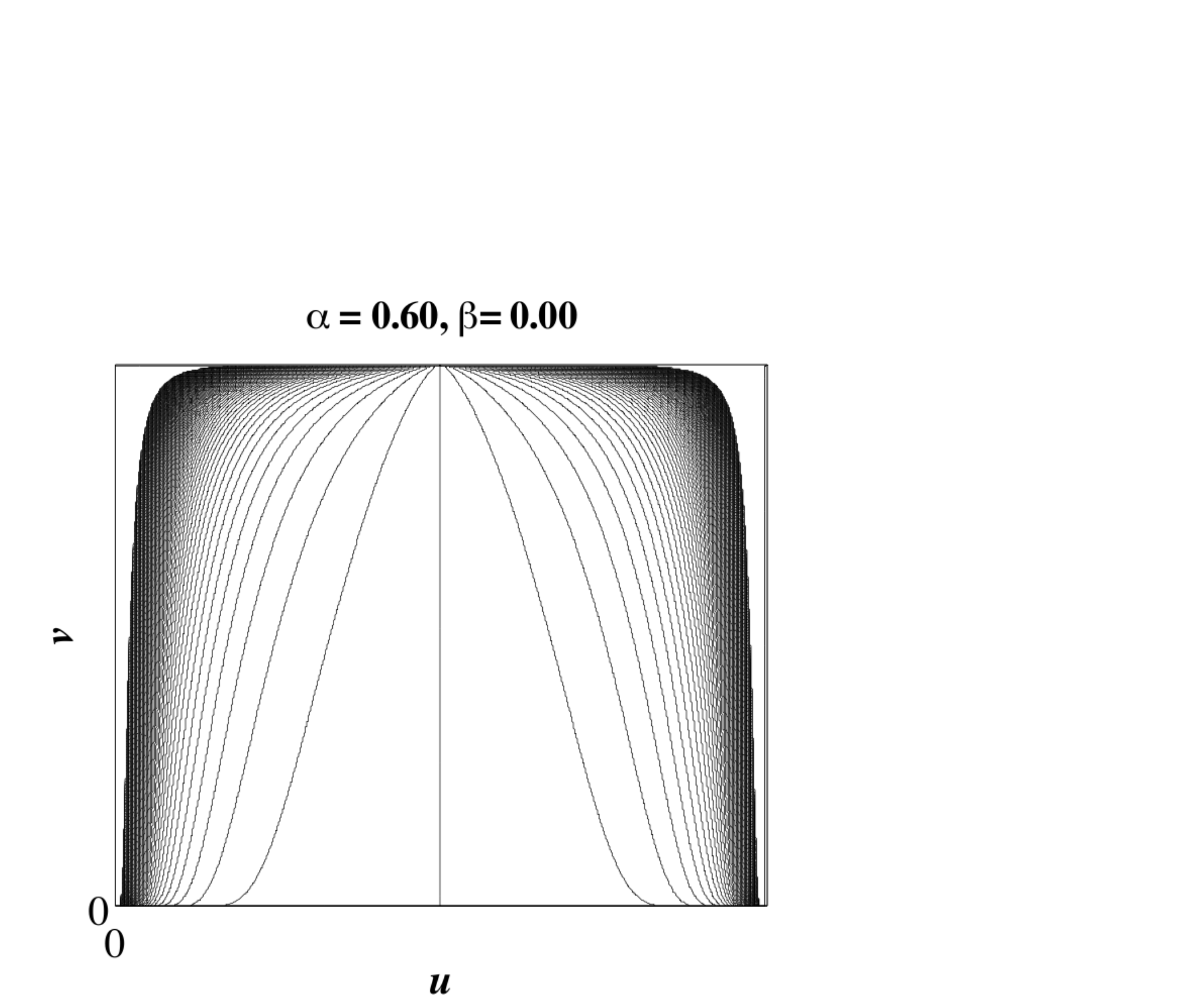}
\begin{picture}(0,0)(0,0)  \put(0.4,3){\scriptsize 1}\put(3.2,0.3){\scriptsize 1}\end{picture}
\includegraphics[clip=true,height=3.5cm,width=3.5cm]{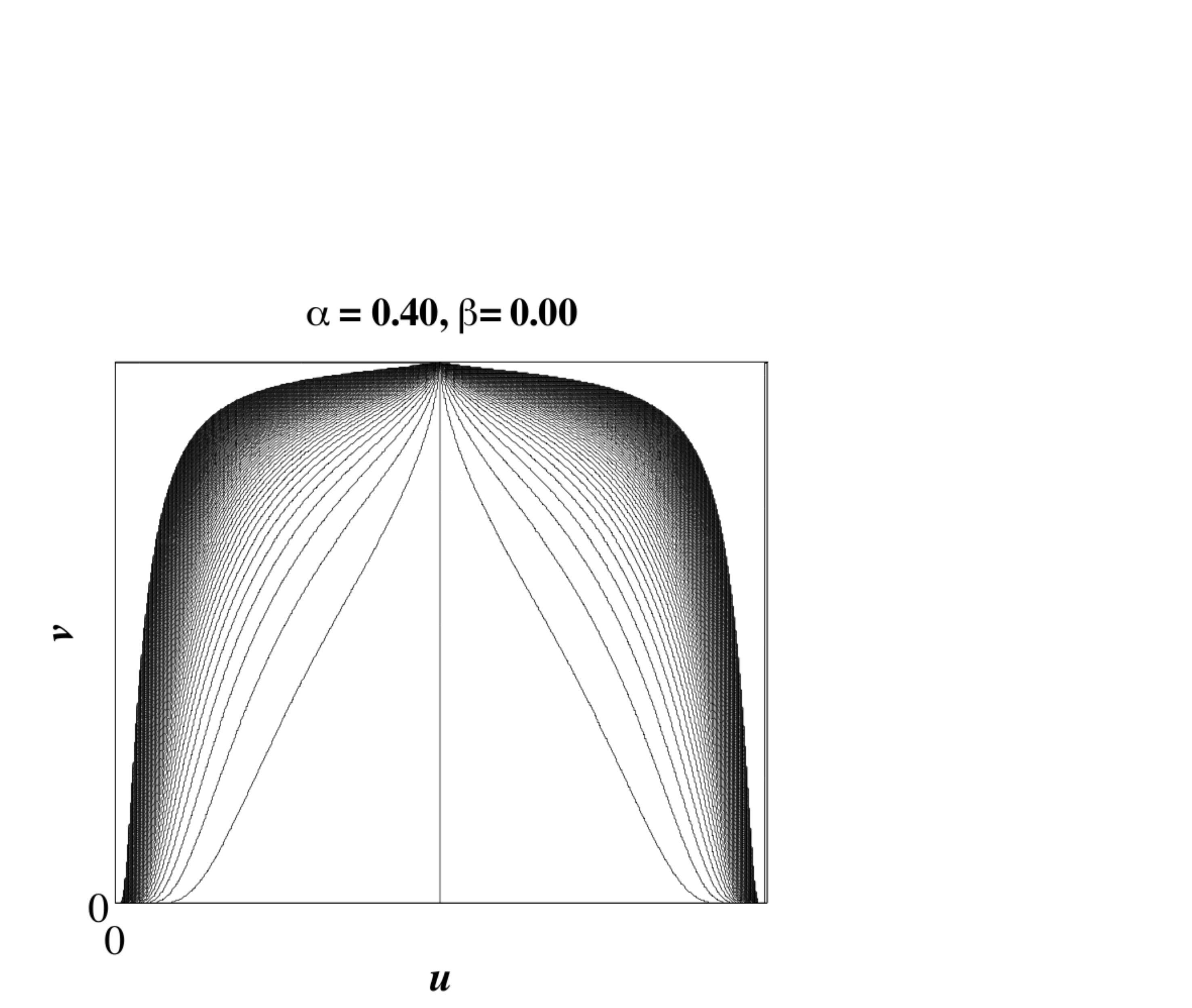}
\begin{picture}(0,0)(0,0)  \put(0.4,3){\scriptsize 1}\put(3.2,0.3){\scriptsize 1}\end{picture}
\includegraphics[clip=true,height=3.5cm,width=3.5cm]{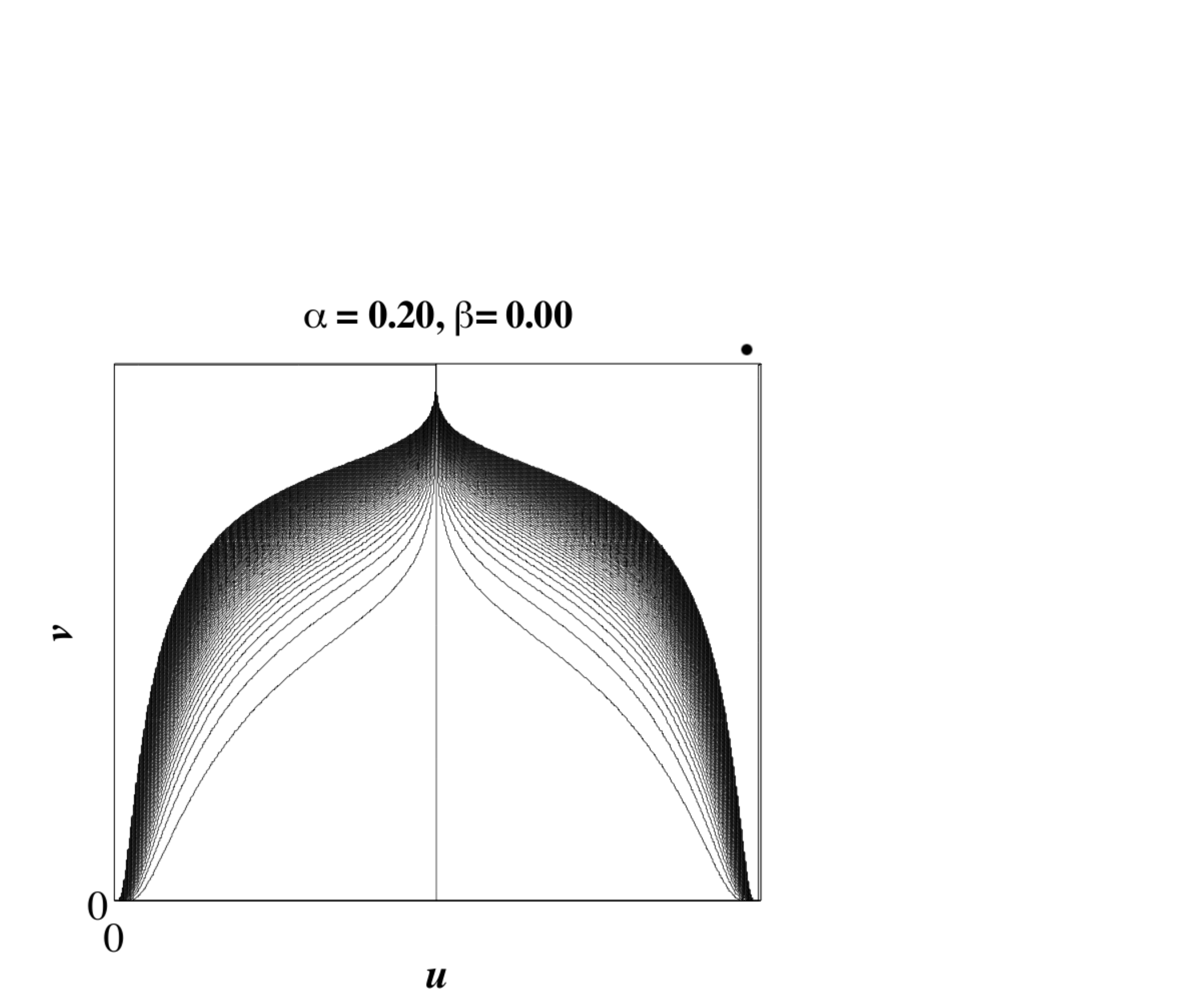}
}
}
\end{center}
\caption{ \label{fig:al12-02}  Isolines of the map $X=F_{\alpha }(u,v)$ with $u,v\in(0,1)$ for decreasing 
values of $\alpha$. The regions with increasingly divergent gradient (upper corners) is not shown
beyond $|x|>600$.  Note that the orientation of the isolines flips over with
decreasing values of $\alpha$ at exactly $\alpha=1$.}
\end{figure}

A transformation method for L\'evy $\alpha$-stable random numbers by Chambers, Mallows and Stuck
has been available for over 30 years~\cite{Chambers1976}. 
Two independent uniform random numbers $U,V \in (0,1)$ are mapped via a transform 
$F_{\alpha\beta}(U,V)$
such that $X = F_{\alpha\beta}(U,V)$ is distributed correctly according to $L_{\alpha\beta}(x)$. 
The general case for $\alpha\ne 1$ is given by
\begin{equation} \label{eq:ChambersAsym}
 X = F_{\alpha\beta}(U,V)    =  \frac{\D\sin(\alpha(\Phi+\Phi_0))}
                                             {\D\cos\Phi}
                                     \left(\frac{\D -\log U \cos\Phi}{\D\cos(\Phi-\alpha(\Phi+\Phi_0))}
                                     \right)^{1-1/\alpha},
\end{equation}
where $\Phi = \pi\left(V-\frac{1}{2}\right)$ and $\Phi_0=\frac{1}{2}\pi\beta\frac{\D 1-|1-\alpha|}{\D \alpha}$,
while for $\alpha=1$
\begin{equation}
        X = F_{\alpha\beta}(U,V) = \left(1+ \frac{2}{\pi}\beta\Phi\right)\tan\Phi - 
        \frac{2\D}{\D\pi}\beta\log\left( \frac{\D -\log U \cos\Phi}{1 + 2\beta\Phi/\pi}\right) .
 \end{equation}
The symmetric case with $\beta=0$ simplifies to
\begin{equation}
 \label{eq:Chamberssym}
  X = F_{\alpha }(U,V) = 
\frac{\sin(\alpha\Phi)}{\cos\Phi}
\left(\frac{ -\log U \cos\Phi }{  \cos((1-\alpha)\Phi) }\right)^{1-1/\alpha} .
\end{equation} 
The variables $X_1,...,X_N$ are stable as well as their normalized sum
\begin{equation}
X=\frac{1}{N^{1/\alpha}}\sum_{i=1}^N X_i.
\end{equation}
This transform representation is a mixture of the form $g(V)W^{1-1/\alpha}$
where $g(V)$ is a real valued 
random number and $W$ is exponentially distributed.  Figs.~\ref{fig:mapsbet=0}
and~\ref{fig:al130be-06} show symmetric and asymmetric examples of
the mapping of the random number plane $(U,V)$ to ``quantiles'' of the
probability density via the map $X=F_{\alpha\beta}(U,V)$.  Colors are used
to designate the respective regions $x_1<X<x_{i+1}$ separated by isolines
defined by $\dd F_{\alpha\beta}(U,V)=0$. The pictures show isolines as
borders between colors for $x_i=0,\pm 0.5,\pm 1,\pm 1.5,...$ . The colors
in the map and in the respective histogram correspond to each other and all
points $(U,V)$ on the same isoline are mapped onto exactly one unique number.
Fig.~\ref{fig:al12-02} shows the behaviour of the isolines only with further
decreasing $\alpha$. Notable is the analytic Cauchy case $\alpha=1$ whose
inversion formula depends only on one variable. This is expressed by perfectly
vertical isolines.  For values of $\alpha<1$ the overall behaviour turns over
and the slopes change sign in each half of the unit square. The pictures
showing isolines are produced with \textsc{Matlab} 7.4's \texttt{contourf}
function on a grid of size 800$\times$800.

We would like to remark that different solutions are thinkable of how to sample
uniform random 
points in a specific region in the $(U,V)$-plane.
A differential equation for the isolines can be obtained via the implicit 
function theorem by Ulisse Dini~\cite{Dini1880}:
\begin{equation}
0  =   \dd F(u,v) =  \frac{\D \partial F(u,v)}{\D\partial u} \dd u 
             + \frac{\D\partial F(u,v)}{\D\partial v} \dd v\ ;
\end{equation}
rearranging 
\begin{equation}\label{eq:DGL}
\frac{\D \dd v(u)}{\D \dd u} = 
- \left(\frac{\D\partial F(u,v)}{\D\partial v}\right) ^{-1}\frac{\D\partial F(u,v)}{\D\partial u}.
\end{equation}
With an appropriate initial condition this differential equation defines the isoline $v(u)$  
in the coordinate square spanned
by $u,v$. The alternative representation of $u$ as a function of $v$ is equally appropriate from the 
mathematical point of view, but is less convenient in this case for symmetry reasons. 
We skip additional considerations on singularities and limiting behaviour.
For $\alpha=2$  and $\beta=0$ Eq.~(\ref{eq:Chamberssym}) 
reduces to $X = F_{2}(U,V) = 2\sqrt{-\log U}\sin(\pi(V-1/2))$, which is the Box-Muller
method for Gaussian deviates with standard deviation $\sigma=\sqrt{2}$. 
The corresponding map is shown in the upper left
of Fig.~\ref{fig:mapsbet=0}.
The value $x$ in the condition $X>x$ determines the
initial condition for  Eq.~(\ref{eq:DGL}) that determines the isoline, 
i.e.\ $x$ in the condition $X>x$, and {for $\alpha=2$} 
it can be chosen on the boundary of the square  $U,V\in(0,1)$. 
Two other analytic limit cases for $\beta=0$, where $L_\alpha(x)$
can be written in terms of elementary functions, are the Cauchy
distribution, with $\alpha=1$ and $X = F_{1}(U)= \tan(\pi(U-1/2))$, and the L\'evy
distribution, with $\alpha=1/2$ and $X =  F_{1/2}(U,V) = - \tan(\pi(V-1/2)) / (2\log U
\cos(\pi(V-1/2)))$. Note that for values of $\alpha\ne2$ the map $F$ is 
singular in the points $(0,1)$ and $(1,1)$.  In such cases the initial condition 
cannot be chosen on the boundary, which considerably complicates the situation numerically.

Starting from the simplest case, insertion of $F_2(u,v)$ into 
Eq.~(\ref{eq:DGL}) yields
\begin{equation}\label{eq:DGLal1}
 \frac{\D \dd v(u)}{\D \dd u} = \frac{\D\cot(\pi v(u))}{\D 2 \pi u\, \log(u)}.
\end{equation}
Insertion of $F_{\alpha }(u,v)$  into Eq.~(\ref{eq:DGL}) yields
\begin{multline}\label{eq:DGLCMS}
 \frac{\D \dd v(u)}{\D \dd u} = (\alpha -1) \left\{ \frac{\D 1}{\D\pi  u \log (u)} 
                                      \left[\tan\left(\pi  \left(v(u)-\frac{1}{2}\right) 
                                  (1-\alpha )\right) (\alpha -1)^2 \right.  \right.\\
  \left.\left. + \cot (\pi  v(u))-\alpha ^2 \cot \left(\pi 
   \left(v(u)-\frac{1}{2}\right) \alpha \right) \right]
      \right\}^{-1}.
\end{multline}
One way to sample directly and uniformly  from the area under $v(u)$
would be an area-preserving map of a square domain spanned by two uniform
random numbers, e.g.\ $U,V\in(0,1)$, or any other suitable two dimensional
domain onto this area. To our knowledge this solution is not available yet.
Alternatively, the function $v(u)$ can be obtained numerically via integration
or by appropriate algorithms for the generation of isolines. Once data points
for $v(u)$ are obtained, any method that samples uniformly the region $X<x$
or $X>x$ is suitable in principle.
 With this, the generation of a tail variable constitutes in itself a standard
non-uniform variate generation task. It is the initial scenario of sampling
uniformly under a curve, but with the great simplification of a finite support.
However, this is  not the route we propose for three reasons. First,
the numerical solution of Eq.~(\ref{eq:DGLCMS}) is cumbersome.  Second,
the initial condition has to be found \emph{within} the $u$-$v$ square
due to the above mentioned singularities.  The subsequent integration in
two directions must be guaranteed to work unattended and automatically
as a black box with $\alpha$ and $\beta$ as the only parameters.  Third,
the outcome is not \emph{exact} in the sense that  the sampled random
tail variates are distributed with respect to an approximated probability
density function based on the data point representation  of the isoline. As
it will turn out a numerical or analytic representation of the isoline is
not a required piece of information and its calculation can be avoided.
It can also be shown that the isolines are monotonic in $u$ in the
regions $F_{\alpha\beta}(u,v)<0$ and $F_{\alpha\beta}(u,v)>0$ which is a
useful property  in Sec.~\ref{sec:sampling}.  Although the approximation
of density functions is commonly accepted as a reasonable compromise in
several applications we introduce in the next section a simple graphical
method without this disadvantage.

\section{Sampling method and example application}\label{sec:sampling}

\begin{figure}
\begin{center}
  \setlength{\unitlength}{1cm}
  \includegraphics[clip=true,height=7.05cm,width=9cm]{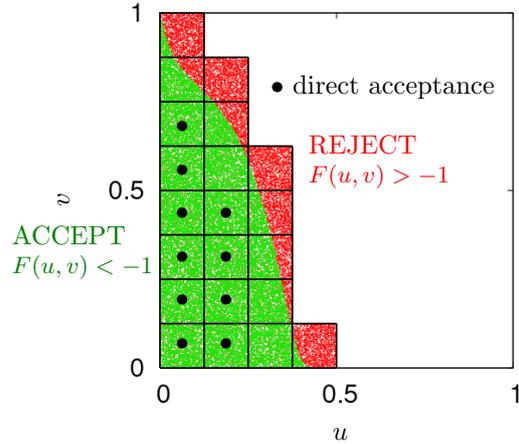} 
  \begin{picture}(0,0)(0,0) 
    \put(-5.2,4){\textcolor{red}{\parbox{2.5cm}{REJECT\\\small $F(u,v)>-1$}}}
    \put(-5.7,5){$\bullet$ direct acceptance}
    \put(-9.1,2.8){\textcolor{darkgreen}{\parbox{3.2cm}{ACCEPT\\\small $F(u,v)<-1$}}}
    \put(-6.95,4.475){$\bullet$}
    \put(-6.95,3.9){$\bullet$}    
    \put(-6.95,3.325){$\bullet$} \put(-6.375,3.325){$\bullet$}
    \put(-6.95,2.75){$\bullet$} \put(-6.375,2.75){$\bullet$}
    \put(-6.95,2.175){$\bullet$} \put(-6.375,2.175){$\bullet$}
    \put(-6.95,1.6){$\bullet$} \put(-6.375,1.6){$\bullet$} 
    \put(-5,0.4){ ${u}$ } 
    \put(-8.5,3.5){\begin{sideways}
            ${v}$ \end{sideways}
                  }
\end{picture}
\end{center}
\caption{ \label{fig:tailexample} (Color online) 
Intuitive, coarsely tiled, example of the tiling in the  $u$-$v$ square for sampling 
symmetric L\'evy $\alpha$-stable random variates
with the condition $X=F(u,v)<-1$ and $\alpha=1.8,\, \beta=0$. 
The tiled area can be sampled efficiently while only points in
the red shaded region are rejected. \tc{darkred}{Tiles with direct acceptance do not 
require the acceptance comparison $X=F(u,v)<-1$.}}
\end{figure}

\begin{figure}
  \setlength{\unitlength}{1cm}
\hspace{-1.6cm}\parbox{1.5\textwidth}{
 \includegraphics[clip=true,height=7cm,width=9cm]{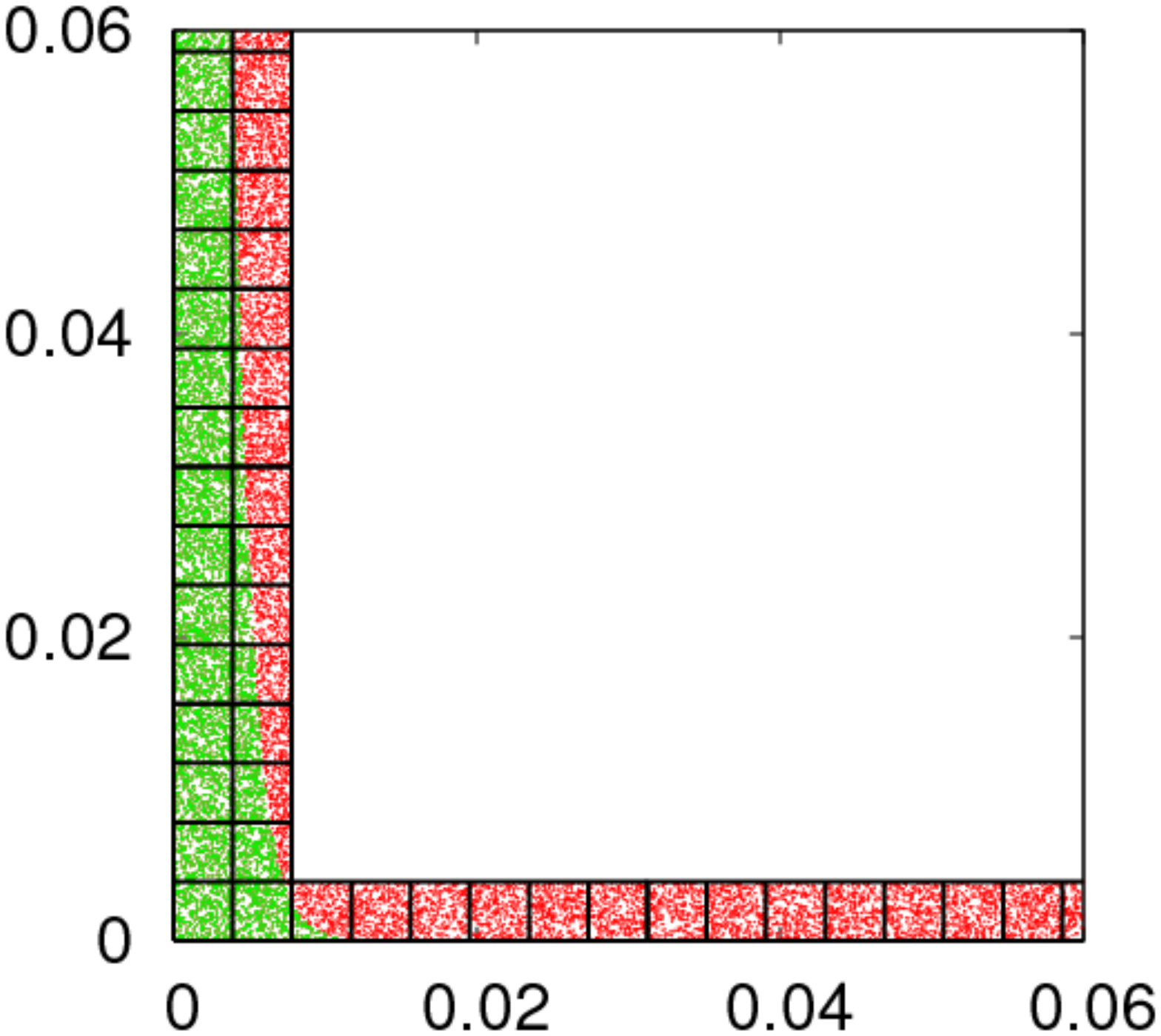}
 \hspace{-1.9cm}\includegraphics[clip=true,height=7cm,width=7.5cm]{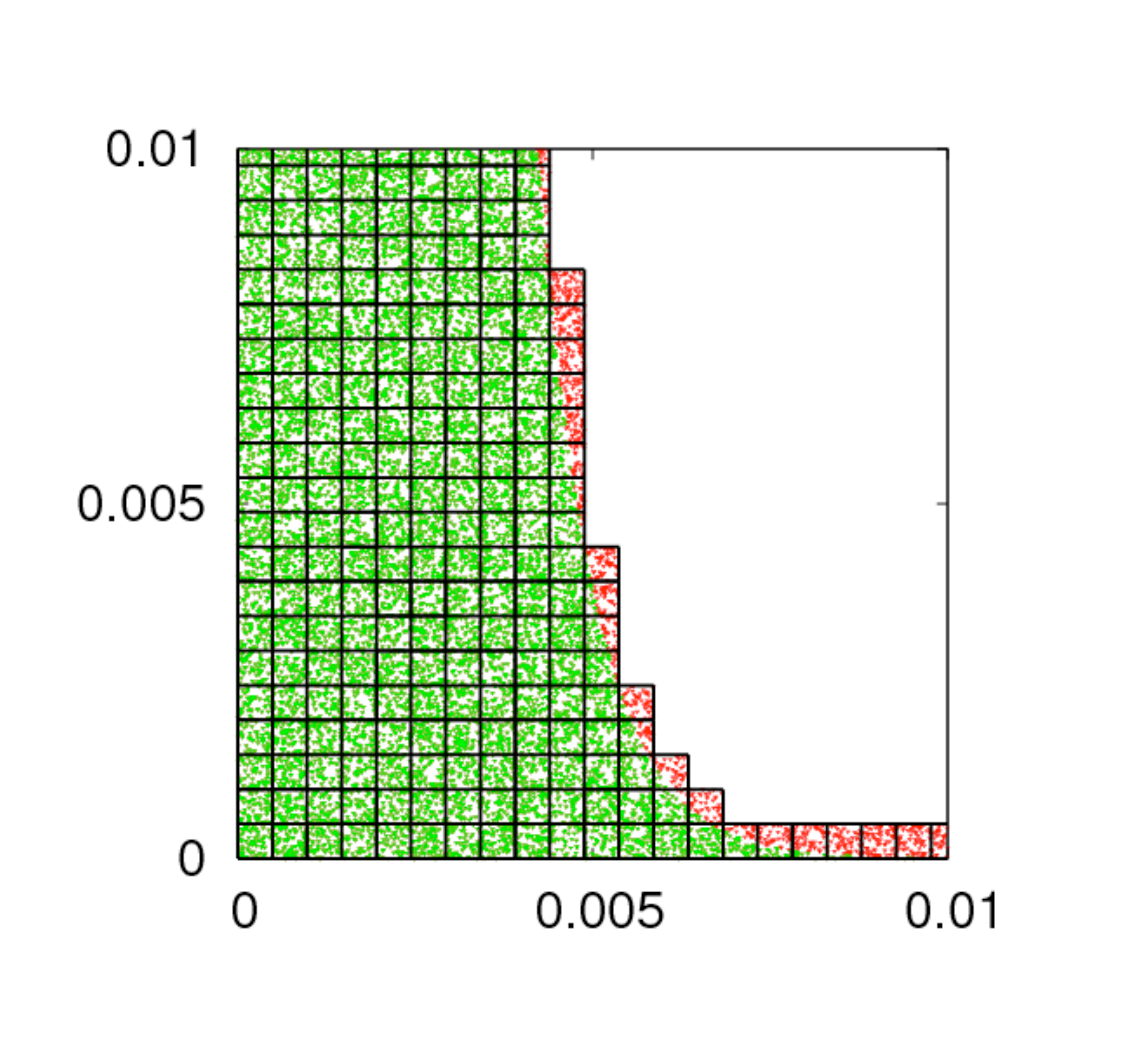} }\\
  \begin{picture}(0,0)(-6,0) 
    \put(-4,3){\textcolor{red}{REJECT}}
    \put(-4,4){\textcolor{darkgreen}{ACCEPT}}
    \put(-3.5,0.5){ ${u}$ }
    \put( 3.4,0.5){ ${u}$ }
    \put( 3,1){\parbox{0cm}{\colorbox{white}{\textcolor{white}{\rule{1cm}{0.3cm}}}}}
    \put( 0.0,3.55){\parbox{0cm}{\begin{sideways}
            \colorbox{white}{${v}$} \end{sideways}\colorbox{white}{\textcolor{white}{\rule{0.5cm}{0.3cm}}}}}
    \put(-7,3.5){\begin{sideways}
            ${v}$ \end{sideways}}
\end{picture}
\caption{ \label{fig:tailsXLT-10} (Color online) Two different tiling refinements 
of the region
corresponding to the condition $X<-12$ which is a  narrow strip along the 
left and bottom of the unit square.
Only the lower left corner is shown on a scale that magnifies the tiling to a visible
size. The row of tiles on the bottom samples a narrow strip below the isoline.
In the right panel the rejection rate is significantly lower. 
It may help intuition that the colored dots are the uniformly distributed pairs  $(u,v)$.}
\end{figure}

We introduce the method using simple intuitive examples.  The production
algorithm relies on the rejection method whose invention dates back
to von Neumann~\cite{vonNeumann1951} and which we do not rehearse here.
Fig.~\ref{fig:tailexample} demonstrates a computationally efficient concept for
uniform sampling in a certain two-dimensional region.  In the first example
we aim at producing L\'evy $\alpha$-stable random variates  with parameters
$\alpha=1.8$, $\beta=0$ and the condition $X<-1$. The map $F_{1.8}(u,v)$
for this choice of parameters is also shown in Fig.~\ref{fig:mapsbet=0}. It
corresponds to a relatively large region in the left part of the square.

We performed a straightforward and simple tiling of this region using square
tiles that can be refined, for example, iteratively maintaining complete
coverage while minimizing the excess area of the tiles that stick out of
the region defined by $F_{1.8}(u,v)<x$.  Uniform sampling of the tiled area
accepting all $X=F_{1.8}(u,v)<x$ and
 rejection of all other samples achieves the desired tail sampling.
The size of the tiles can be chosen to achieve an arbitrarily low rejection
rate.  In the example shown in Fig.~\ref{fig:tailexample} the tiling is
refined only moderately to convey the situation. For tiles that lie completely
underneath the isoline the test $F_{1.8}(u,v)<x$ must not be executed.
With dense tiling this comparison is therefore hardly needed and indeed must
be avoided to yield a speed-up with respect to the transformation method.
The setup and production loop of random variates follows the steps:

\noindent
\begin{minipage}{\textwidth}
\noindent
\textbf{Algorithm}
Input: $x\in\mathbb{R}$.
\begin{description}
\item[0:] Setup:\\ 
         Tiling of the region $F(u,v)<x$ using a method of choice.\\
         Label tiles with an integer index.\\
         Label tiles that are intersected by the isoline $F(u,v)=x$.
\item[1:] Draw a random integer tile index with uniform probability.
\item[2:] Draw a random coordinate $(u,v)$ with uniform probability within this tile.
\item[3:] Test if the tile is intersected by the isoline (table look-up).\\
         If yes, go to 4. If no, accept $X=F(u,v)$ and go to 1 (direct acceptance).
\item[4:] Test if $(u,v)$ satisfies $X=F(u,v)<x$. \\
         If yes, accept $X$, otherwise reject  and go to 1.
\end{description}
\end{minipage}
Note that for monotonic isolines the position of a tile with respect to an
isoline, i.e.\ whether underneath, above or intersected, can be determined
by evaluating the map for at most two corners.  Step 4 is unlikely to be
carried out if the coverage is dense,  giving nearly a zero rejection rate.
Overall, this procedure is efficient in setup and production for a sufficiently
dense tiling.  Furthermore, with small modifications of the above acceptance
and rejection conditions in the pseudo code, the tiling and production of
random numbers on a finite interval $X\in[x_1,x_2]$ is  geometrically and
algorithmically equivalent to generating numbers from the tail.  This requires
the tiling of a region in the $u$-$v$ square between two isolines with the
condition  $x_1<X<x_2$.

Fig.~\ref{fig:tailsXLT-10} shows the map for the left tail regions of the
$u$-$v$ square with $X<-12$, which is more realistic for the purpose of
tail sampling.  This condition corresponds to sampling a narrow strip at
the bottom and left sides of the unit square. The figure only shows the
corner at the origin. The bottom strip of tiles samples an extremely narrow
strip than is not visible on this scale.  The iterative tile refinement in
the setup stage is acceptably fast, below a second in our non-optimized
code, down to the level on the right panel of Fig.~\ref{fig:tailsXLT-10}
to achieve a rejection rate below 1\%.  Different values of $\alpha>0.1$ as
well as not too extreme values of $x$ have no significant influence on the
setup performance achieving a rejection rate of 1\%.  Note that the speed of
random number production is independent of the number of tiles.  In our case it
amounts to 2.3 million tail variates per second on a PC with a 2.4 GHz Intel
Pentium 4 processor  using the GNU C++ compiler version 3.2.2 
Linux and optimization level -O3. As the uniform random number generator we
used the XOR shift SHR3 by Marsaglia~\cite{Marsaglia2000}.  The colouring
of the acceptance and rejection regions in Figs.~\ref{fig:tailexample}
and~\ref{fig:tailsXLT-10} are produced by green and red coloured dots
representing the random uniform coordinates $(u,v)$.

We would like to stress that the method of tiling as well as the form of the tiles
is in principle arbitrary. Equal size and shape is computationally advantageous,
but this issue is not the focus of the present work. 
Of course any  tiling technique that produces a similar 
result is suitable, using either 
square or rectangular tiles. However, the choice of square
equal tiles is algorithmically very simple and likely to outrun an adaptive scheme
with more complex shapes in setup and also production.
The iterative tiling refinement, as performed in the above examples,
is robust and fast also for large values of  $|x|$.
The rejection scheme is in principle similar to the Ziggurat implementation of 
Ref.~\cite{Marsaglia2000}. It also needs the setup of
a data structure that  covers  a region by equal area rectangles.
The details of the tiling method that 
is more general and applicable to random number production directly via the
probability density are described in separate work~\cite{Fulger2008b}.

\section{The Mittag-Leffler probability distribution}\label{sec:ML}

\begin{figure}
\begin{center}
\scalebox{1.26}{
\hspace{-12mm}\parbox{\textwidth}{
\begin{center}
 \setlength{\unitlength}{1cm}
\includegraphics[clip=true,height=3.9cm,width=3.9cm]{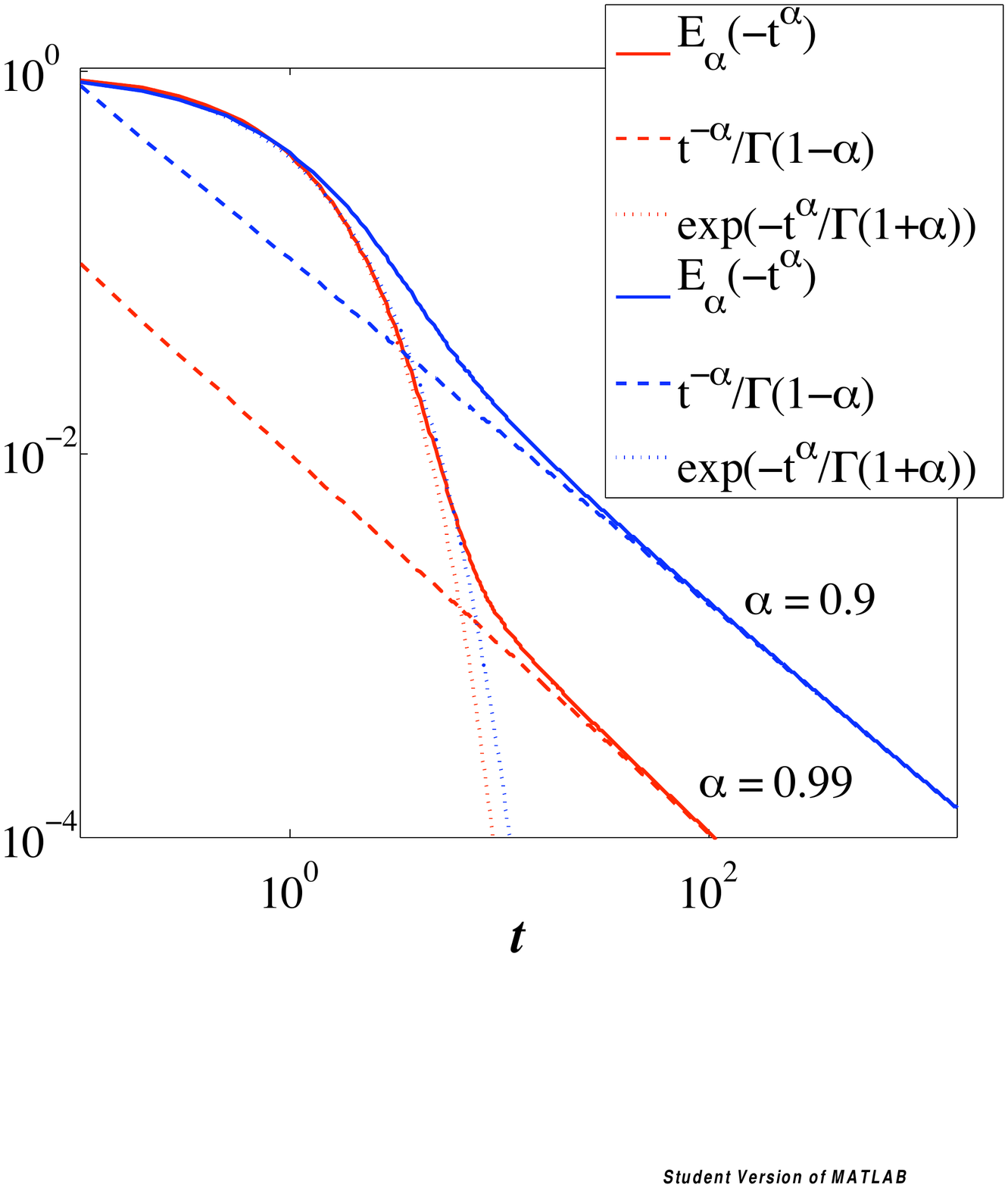}
\begin{picture}(0,0)(0,0)
\put(0.4,3){\scriptsize 1}\put(3.2,0.3){\scriptsize 1}
\end{picture}
\includegraphics[clip=true,height=3.5cm,width=3.5cm]{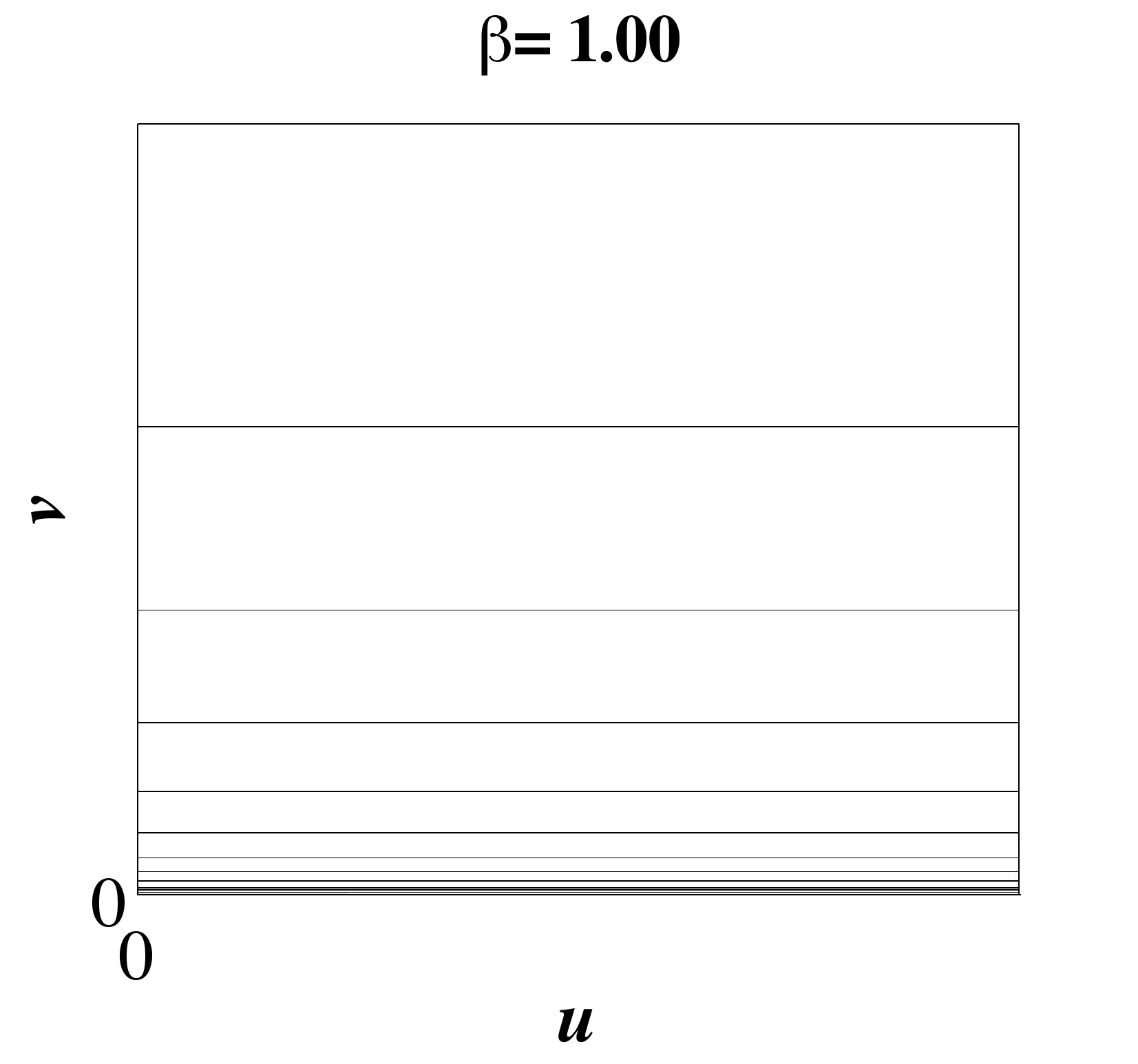}
\begin{picture}(0,0)(0,0)
     \put(-2.45,3.3){\scriptsize \colorbox{white}{$\alpha=1.00$}}
     \put(0.4,3){\scriptsize 1}\put(3.2,0.3){\scriptsize 1}
\end{picture}
\includegraphics[clip=true,height=3.5cm,width=3.5cm]{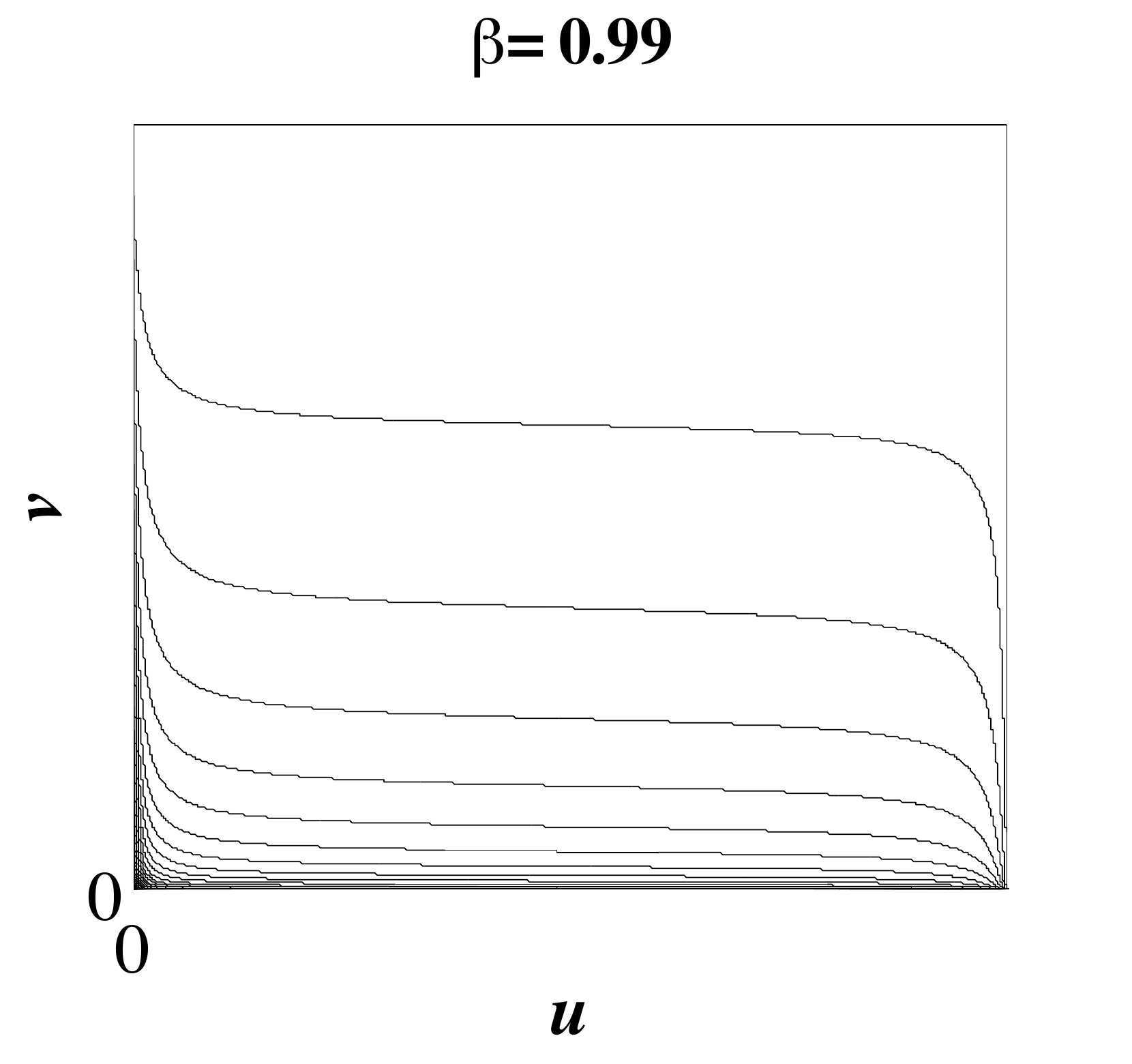}
\begin{picture}(0,0)(0,0)
      \put(-2.45,3.3){\scriptsize \colorbox{white}{$\alpha=0.99$}}
\end{picture}
\includegraphics[clip=true,height=3.5cm,width=3.5cm]{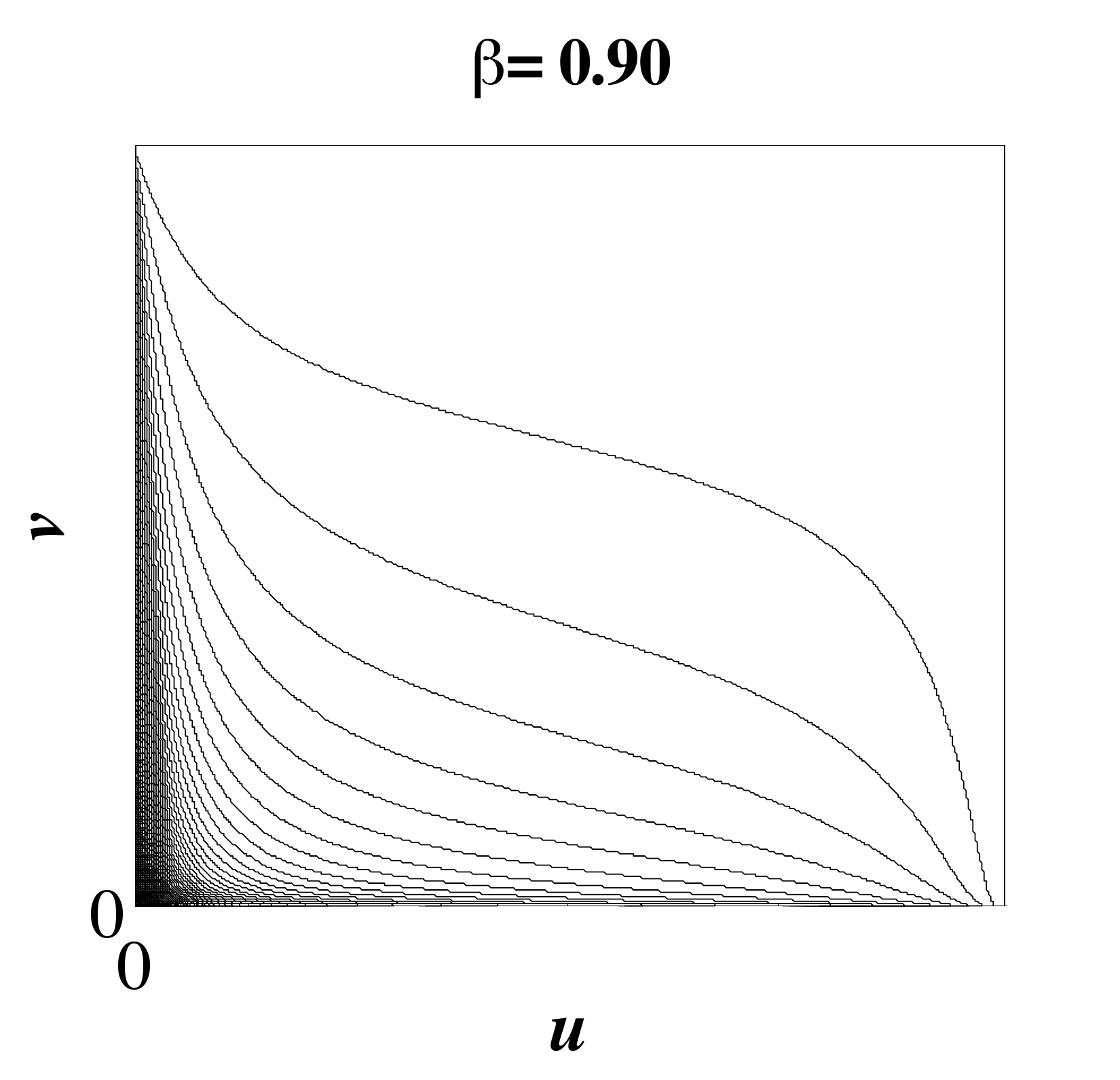}
\begin{picture}(0,0)(0,0)
 \put(-2.45,3.3){\scriptsize \colorbox{white}{$\alpha=0.90$}}
 \put(-3.34,3){\scriptsize 1}\put(-0.5,0.3){\scriptsize 1}
 \put(0.4,3){\scriptsize 1}\put(3.2,0.3){\scriptsize 1}
\end{picture}
\includegraphics[clip=true,height=3.5cm,width=3.5cm]{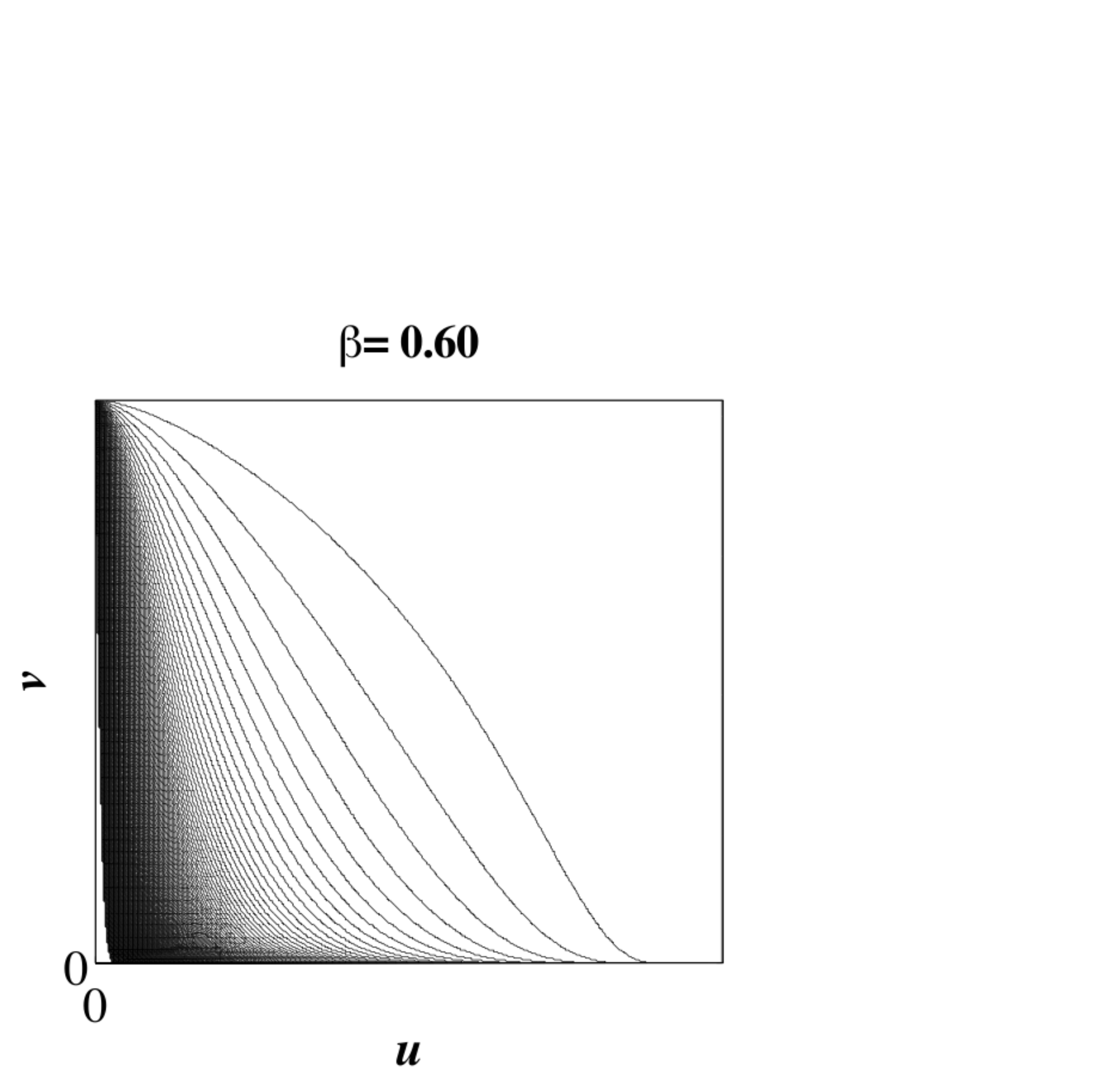}
\begin{picture}(0,0)(0,0)
  \put(-2.45,3.3){\scriptsize \colorbox{white}{$\alpha=0.60$}}
 \put(0.4,3){\scriptsize 1}\put(3.2,0.3){\scriptsize 1}
\end{picture}
\includegraphics[clip=true,height=3.5cm,width=3.5cm]{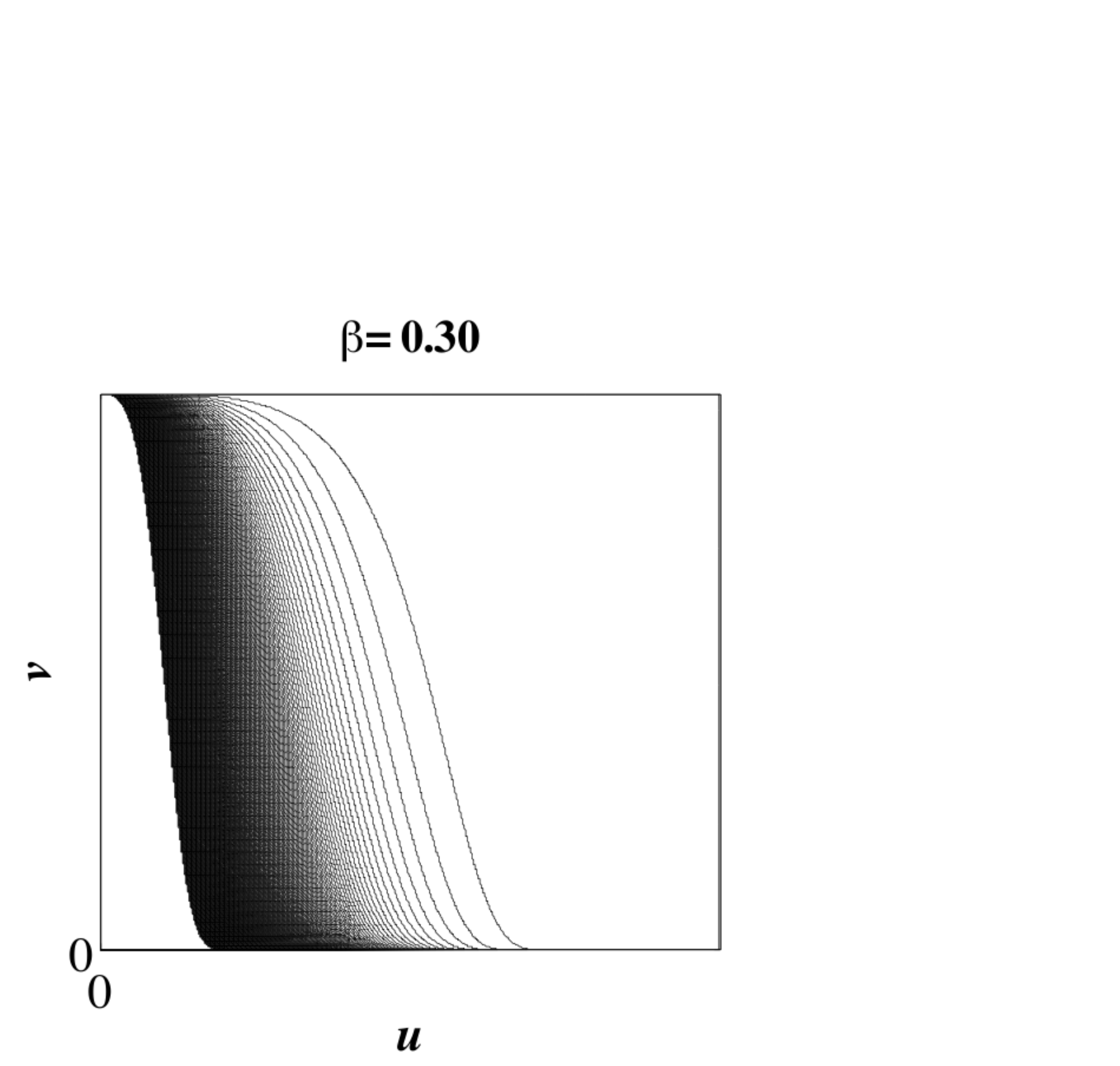}
\begin{picture}(0,0)(0,0)
  \put(-2.45,3.3){\scriptsize \colorbox{white}{$\alpha=0.30$}}
\end{picture}
\includegraphics[clip=true,height=3.7cm,width=6.8cm]{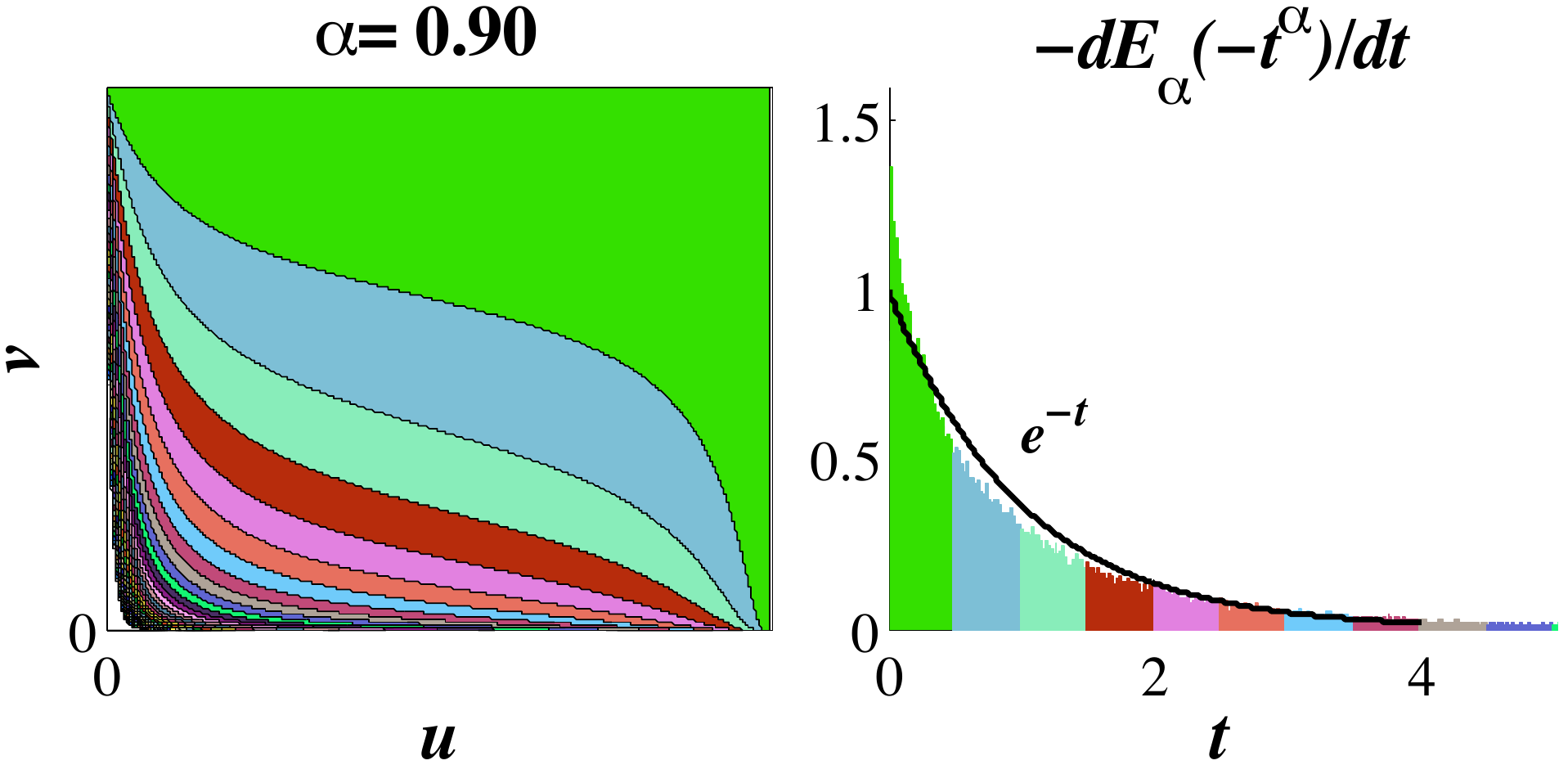}
\begin{picture}(0,0)(0,0) \put(-6.7,3.17){\scriptsize 1}\put(-3.65,0.35){\scriptsize 1}
\put(-5.65,3.4){\colorbox{white}{$\alpha=0.9$}} 
\end{picture}
\end{center}
}
}
\end{center}
\caption{ \label{fig:MLmaps}  The Mittag-Leffler function $E_\alpha(-t^\alpha)$
in a log-log plot (top left) and the transformation map $X=M_\alpha(U,V)$,
Eq.~(\ref{eq:Kozubowski}), in terms of isolines for five values of $\alpha$.
The case $\alpha=1$ corresponds to the standard exponential function. The
regions on the left side of the maps is not shown beyond $t>600$ due to an
increasingly divergent gradient. The two plots at the bottom repeat the case
with $\alpha=0.9$ using colors and showing the corresponding histogram.}
\end{figure}

Our second example density is less know in scientific applications, even less
so its transform. The Mittag-Leffler probability distribution appears e.g.\ in
the analytic solution of the time-fractional Fokker-Planck
equation~\cite{Fulger2008,Gorenflo2007,Gorenflo2004,Hilfer1995}.
The generalized Mittag-Leffler function is defined
as~\cite{Gorenflo2002,Hilfer2006}
\begin{equation}
E_{\alpha\beta}(z) = \sum_{n=0}^\infty \frac{z^n}{\Gamma(\alpha n+\beta)}, \quad
z \in \mathbb{C}.
\end{equation}
For our purposes it is sufficient to restrict the example to the one-parameter 
Mittag-Leffler function which plays an important role in the stochastic
solution of the time-fractional diffusion equation. The series representation
is
\begin{equation}
\label{eq:MLpowerseries}
E_\alpha(z) = \sum_{n=0}^\infty \frac{z^n}{\Gamma(\alpha n+1)}, \quad
z \in \mathbb{C},
\end{equation}
leading to a pointwise representation on a finite interval. The Mittag-Leffler
function with argument  $z=-t^\alpha,\ t \in \mathbb{R}$ reduces to a standard
exponential decay, $e^{-t}$, with $\alpha = 1$; when $0 < \alpha < 1$, the
Mittag-Leffler function is approximated for small values of $t$ by a stretched
exponential decay (Weibull function),
$\exp(-t^\alpha/\Gamma(1+\alpha))$ and for large values of $t$ by a power law,
$t^{-\alpha}/\Gamma(1-\alpha)$, see Fig.~\ref{fig:MLmaps}, top left plot.
There is increasing evidence for physical phenomena \cite{Clauset2007,Mega2003,
Shlesinger1993,Ward1998} and human activities \cite{Barabasi2005,Raberto2002,
Scalas2004b} that do not follow neither exponential nor, equivalently,
Poissonian statistics. The Mittag-Leffler distribution is an example of
power-law distributed waiting times. They arise as the natural survival
probability leading to time-fractional diffusion equations.  

Eq.~(\ref{eq:MLpowerseries}) is the complementary cumulative distribution
function, also called survival function; the proability density is
\begin{equation}
\label{eq:MLdens}
-\frac{\dd}{\dd t}E_\alpha(-t^\alpha). 
\end{equation}
In past applications Mittag-Leffler random numbers were produced by rejection
through a pointwise representation via Eq.~(\ref{eq:MLpowerseries}), which is
inefficient due to the slow convergence of the series. In some cases concepts
to avoid Mittag-Leffler random numbers were presented~\cite{Magdziarz2007b,
Magdziarz2007a} due to the difficulty of their production. In this context it
had not been recognized immediately that transformation formulas analogous to
Eq.~(\ref{eq:Chamberssym}) are available~\cite{Devroye1996,Germano2006,
Jayakumar2003,Kozubowski1998,Kozubowski2000,Kozubowski2001,Kozubowski1999,
Pakes1998}. The most convenient expression is due to Kozubowski and
Rachev~\cite{Kozubowski1999}:
\begin{equation}
\label{eq:Kozubowski}
T = M_\alpha(U,V) = \log U
\left(\frac{\sin(\alpha\pi)}{\tan(\alpha\pi V)}-\cos(\alpha\pi)\right)^{1/\alpha},
\end{equation}
where $U, V \in (0,1)$ are independent uniform random numbers and $T$ is a
Mittag-Leffler random number. For $\alpha=1$, Eq.~(\ref{eq:Kozubowski}) reduces
to the transform for the exponential distribution: $M_1(U,V) = \log U$. 
Fig.~\ref{fig:MLmaps} shows the map $M_\alpha(U,V)$ of the transform
representation Eq.~(\ref{eq:Kozubowski}) as borders between intervals
corresponding to $ t=0,\pm 0.5,\pm 1,\pm 1.5,...$ The exponential case with
$\alpha=1$ depends on only one random variable in the $u$-$v$ square which is
expressed by perfectly horizontal isolines. For $\alpha<1$ the left and right
edges develop singularities. It is not recommended to use
Eq.~(\ref{eq:MLpowerseries}) and summation of many terms for the computation of
$E_\alpha(-t^\alpha)$. A more elegant and accurate method is presented in
Ref.~\cite{Gorenflo2002,Podlubny2005,Hilfer2006}. For the generation of random
numbers we use the implementation in Ref.~\cite{Germano2008}.

\clearpage

\section{Summary and conclusion}

We have demonstrated some properties of the Chambers-Mallows-Stuck and
Kozubowski-Rachev transformation maps exemplifying the production of random
numbers with the former. The interpretation as a two-dimensional map from the
unit square to the real numbers allows to associate arbitrary intervals on the
support of the density with well defined finite regions of the map domain.
The uniform sampling of such regions produces directly random numbers exactly
within the respective intervals. We have also introduced an efficient concept
for the automatic setup of a random number generator that makes use of this
property. The resulting generator can in principle produce random numbers in 
intervals which can be disconnected and any combination of the kind $(-\infty,
x_1] \cup [x_2,x_3]\cup \ldots\cup [x_{n},+\infty),\, x_i \in\mathbb{R}$.
Most importantly, the sampling of tails as shown here can be used as a tail
handling method in fast implementations of random number generators for which
a candidate is the Ziggurat method implementation that was proven to greatly
outrun simple inversion methods. The present work opens the route for the
speedup of many known random number generators that rely on transform
representations.

\bibliographystyle{amsplain}
\bibliography{journals,paper}

\end{document}